\documentclass[aps,prd,reprint,onecolumn,groupedaddress]{revtex4-2}

\usepackage{graphicx}
\usepackage[caption=false]{subfig}
\usepackage{xspace}
\usepackage[dvipsnames]{xcolor}
\usepackage{comment}
\usepackage{enumerate}
\usepackage{hyperref}
\usepackage{amsmath}
\usepackage{amssymb}

\newcommand{\argmaxNL}{\mathop{\mathrm{arg~max}}}
\newcommand{\argminNL}{\mathop{\mathrm{arg~min}}}

\newcommand{\mmax}{\mathrm{max}}
\newcommand{\mmin}{\mathrm{min}}

\newcommand{\dblhat}[1]{\hat{\hat{#1}}}
\newlength{\dhatheight}

\makeatletter
\newcommand*{\coloneq}{\mathrel{\rlap{%
                     \raisebox{0.3ex}{$\m@th\cdot$}}%
                     \raisebox{-0.3ex}{$\m@th\cdot$}}%
                     =}
\makeatother
\makeatletter
\newcommand*{\eqcolon}{=%
                     \mathrel{\rlap{%
                     \raisebox{0.3ex}{$\m@th\cdot$}}%
                     \raisebox{-0.3ex}{$\m@th\cdot$}}}
\makeatother

\newcommand{\EE}{\mathbb E}
\newcommand{\Var}{\mathrm{\mathop{Var}}}

\newcommand{\epsilonValue}{0.3}
\newcommand{\DchimaxValue}{32}

\begin{document}

\title{An importance sampling method for Feldman-Cousins confidence intervals}

\author{Lukas Berns}
\email[]{lukasb@epx.phys.tohoku.ac.jp}
\affiliation{Tohoku University, Japan}

\date{\today}

\begin{abstract}
	In various high-energy physics contexts, such as neutrino-oscillation experiments, 
	several assumptions underlying the typical asymptotic confidence interval construction are violated,
	such that one has to resort to computationally expensive methods like the Feldman-Cousins method
	for obtaining confidence intervals with proper statistical coverage.
	By construction, the computation of intervals at high confidence levels requires fitting millions or billions of pseudo-experiments, while wasting most of the computational cost on overly precise intervals at low confidence levels.
	In this work, a simple importance sampling method is introduced which reuses pseudo-experiments produced for all tested parameter values in a single mixture distribution.
	This results in a significant error reduction on the estimated critical values, especially at high confidence levels,
	and simultaneously yields a correct interpolation of these critical values between the parameter values at which the pseudo-experiments were produced.
	The theoretically calculated performance is demonstrated numerically using a simple example from the analysis of neutrino oscillations.
	The relationship to similar techniques applied in statistical mechanics and $p$-value computations is discussed.
\end{abstract}

\maketitle

\section{Introduction}

An essential part of any experiment is the statistical analysis to extract information about the model parameters, such as physics constants, from the measurement outcome. As measurements inherently include statistical fluctuations, one often reports these constraints in the form of confidence intervals (or confidence regions in higher dimensions). These are intervals over the parameter space calculated from the observed data, which are constructed in such a way that for any true value of the parameters, at least a pre-defined percentage of the possible experimental outcomes would produce an interval that covers the true parameter value. The pre-defined percentage over possible experimental outcomes is called the confidence level (CL).

For the rest of this paper we shall use the following notation: $x$ denotes the experimental outcome, which can be a vector of many observations within the single experiment. $\theta$ denotes the model parameters, which can contain one or higher dimensional continuous degrees of freedom, and may contain discrete degrees of freedom as well. $p(x \mid \theta)$ denotes the probability distribution function for the experimental outcomes given some model parameters. $p(x \mid \theta)$ seen as a function of $\theta$ for a given experimental outcome is called the likelihood function and denoted $L(\theta \mid x) \coloneq p(x \mid \theta)$. The parameter value for which the likelihood is maximized is denoted $\hat \theta(x) \coloneq \argmaxNL_\theta L(\theta \mid x)$, and the difference of the log-likelihood at some parameter value to the maximum likelihood is denoted as $\Delta \chi^2(\theta \mid x) \coloneq -2 \log L(\theta \mid x) / L(\hat\theta(x) \mid x)$. The confidence level is denoted $1 - \alpha$.

In many cases a useful theorem by Wilks~\cite{Wilks:1938dza} can be applied, which greatly simplifies the construction of such confidence intervals. The theorem says that
in the asymptotic limit, $\Delta \chi^2(\theta \mid x)$ evaluated at the true parameter value is distributed as a chi-squared distribution with $k$ degrees of freedom, where $k$ is the dimension of the parameter space $\theta$, which has to be continuous. The theorem holds under suitable conditions which ensure that a maximum likelihood value can be found in the neighborhood of the true parameter value with a quadratic Taylor expansion of the likelihood. Given this asymptotic distribution, one can thus construct a confidence interval by all values of $\theta$ that satisfy $\Delta \chi^2(\theta \mid x) \le \Delta \chi^2_c$, where the critical value $\Delta \chi^2_c$ is easily computed from the quantile function of the chi-squared distribution.

Due to the necessary assumptions, confidence intervals based on Wilks' theorem are not suitable if the number of observations is small, or the parameter space is unsuitable because of physical boundaries (such as $\theta \ge 0$), 
discrete degrees of freedom, or periodicities that cannot be captured by the quadratic expansion. Neutrino oscillation experiments for example suffer from all of these deficiencies, for which we will present an example later.
In this situation, one has to resort to actually producing ensembles of pseudo-experiments for selected parameter values to study the distribution of a suitable statistic to be used for the construction of the confidence interval.

A commonly used method is the Feldman-Cousins (FC) method~\cite{Feldman:1997qc}, where for each pseudo-experiment $x'$ generated assuming a true value $\theta_t$, the $\Delta \chi^2(\theta_t \mid x')$ value at the true parameter value is computed to obtain its distribution. Then the critical value $\Delta \chi^2_c$ is obtained by the empirical $1 - \alpha$ percentile of this distribution. Since the distribution of $\Delta \chi^2(\theta_t \mid x')$ will in general be different for each true parameter value, the critical values are now a function of the true value at which they are computed, which we denote as $\Delta \chi^2_c(\theta_t)$. Finally, the confidence interval for the actually observed data $x$ is constructed by choosing $\Delta \chi^2(\theta \mid x) \le \Delta \chi^2_c(\theta)$. In practice, it is only possible to compute $\Delta \chi^2_c(\theta)$ at selected parameter values, which need to be interpolated, for example linearly, in order to compute the confidence intervals.

The Feldman-Cousins method is very inefficient for obtaining high-CL intervals, because by definition, only a small fraction of pseudo-experiments contribute to the quantile computation. For example, in particle physics the threshold for ``discovery'' is commonly chosen at $\alpha = 5.7 \times 10^{-7}$ (the ``$5\sigma$'' threshold), in which case only one in 1.7~million pseudo-experiments would (by definition) have a $\Delta \chi^2(\theta_t \mid x')$ value larger than the critical value. As a result, one easily ends up with millions of pseudo-experiments to be fitted in order to obtain the necessary critical values, while simultaneously ``wasting'' most of this computation time for over-precise critical values at lower CL. In practice, FC confidence intervals are often computed only up to $2\sigma$ ($\alpha = 4.6 \times 10^{-2}$) or $3\sigma$ CL ($\alpha = 2.7 \times 10^{-3}$) for such reasons.

In this work, we show that it is actually extremely easy to introduce an alternative sampling distribution that generates high-CL pseudo-experiments much more frequently: one simply reuses the pseudo-experiments generated at the values of the parameters in the form of a mixture distribution. By appropriate reweighting, this results in an exponential reduction in the errors on critical values for high CL. The method also introduces a method for correctly interpolating the critical values between the subset of true parameter values, thus removing the need of naive
interpolation methods that are commonly employed.

The paper is organized as follows. First, we review the conventional FC method.
Next we define the new method, deriving it from a discussion of an ideal importance sampling distribution.
Bounds for the importance sampling weights are calculated, which are used to calculate the reduction of errors on the estimated critical values compared to the conventional FC method. The ability to interpolate critical values and the calculation of errors and other other diagnostics are discussed.
Next, a toy example from the analysis of neutrino oscillations is used to compare the two computation methods and the improvement is checked against the theoretical upper bounds from the previous section.
Finally, we discuss the relationship  to similar techniques in statistical mechanics and $p$-value calculations, the relationship to Bayesian marginalized likelihoods, and the limit of applicability in the presence of nuisance parameters.

\section{Critical values in the conventional Feldman-Cousins method}

To prepare the notation, we briefly review the computation of critical values in the conventional Feldman-Cousins method.
First, we make a choice of $S$ points in the parameter space, which we denote $\theta_s$ with $s$ going from $1$ to $S$. At each $\theta_s$, we now generate an ensemble of $n_\mathrm{exp}$ pseudo-experiments $\{x\}_s$ by sampling from $p(x \mid \theta_s)$. While all pseudo-experiments are assumed to live in the same space, the $s$ suffix on the curly brackets representing the ensemble keeps track of the distribution that generated the experiments. For each pseudo-experiment $x \in \{x\}_s$, we now compute $\Delta \chi^2(\theta_s \mid x)$ and find the $1-\alpha$ quantile $\Delta \chi^2_{c,s}$ through any suitable estimator. For example, one may simply sort the $\Delta \chi^2(\theta_s \mid x)$ values and take the $\lfloor \alpha \times n_\mathrm{exp} \rfloor$ largest value as $\Delta \chi^2_{c,s}$, in which case we have
\begin{equation}
	\sum_{x \in \{x\}_s} I\bigl(\Delta \chi^2(\theta_s \mid x) \ge \Delta \chi^2_{c,s}\bigr) = \lfloor \alpha \times n_\mathrm{exp} \rfloor
	.
\end{equation}
Here, $I(\cdot)$ is the indicator function returning $1$ if the logical statement in the parentheses is true, and $0$ otherwise. $\lfloor \cdot \rfloor$ denotes the floor function.

Finally, the critical value function $\Delta \chi^2_{c}(\theta)$ is obtained by some interpolation scheme. For example, one may set $\Delta \chi^2_{c}(\theta_s) \coloneq \Delta \chi^2_{c,s}$ and linearly interpolate for any $\theta$ values in between.
To reduce the interpolation error, one typically has to either manually or automatically~\cite{Li:2018ewc} adjust the choice of sampling parameter values $\{\theta\}_S$ in an iterative scheme.

The asymptotic variance on the critical values is proportional to the binomial error $ \alpha (1-\alpha) / n_\mathrm{exp}$, so high-CL ($\alpha \ll 1$) generally means that one needs $n_\mathrm{exp} \gg 1/\alpha$ for reliable critical values. Since the whole process is repeated for all $S$ points in the parameter space, the total number of generated (and fitted) pseudo-experiments is $S \times n_\mathrm{exp} \gg S/\alpha$.

\section{The mixture Feldman-Cousins method}

\subsection{Definition}

Our new method, which we shall refer to as the ``mixture Feldman-Cousins'' method, differs from the conventional method mainly in the reuse of \emph{all} generated pseudo-experiments for the critical-value computation of \emph{each} target parameter space point $\theta_t$ with an additional weight 
\begin{equation}
	w(x \mid \theta_t) \coloneq \frac{p(x \mid \theta_t)}{\frac{1}{S} \sum_{s = 1}^S p(x \mid \theta_s)}
	= \frac{1}{\frac{1}{S} \sum_{s = 1}^S \exp[-\frac{1}{2}\{ \Delta\chi^2(\theta_s \mid x) - \Delta\chi^2(\theta_t \mid x) \}]}
	\label{eq:mixture:def:weight}
	.
\end{equation}
The value in the denominator is the sampling probability distribution of $x \in \{x\}_\mathrm{mix} \coloneq \bigcup_{s=1}^S \{x\}_s$, which is the mixture distribution of $p(x \mid \theta_s)$ for all $\theta_s$ values.
Since the weights are based on the sampling probabilities which are nothing but the likelihood function,
they are computable using the same procedure that calculates the $\Delta \chi^2(\theta \mid x)$ for each pseudo-experiment.
Due to taking the difference of two $\Delta\chi^2$ values, the contribution from the minimum $\chi^2$ at $\hat\theta(x)$, as well as any $\theta$-independent offsets (e.g.\ the $n!$ factor in the Poisson likelihood) vanish in the denominator and hence do not need to be known accurately.
While in the conventional method one only needs to compute $\Delta \chi^2(\theta_s \mid x)$ for the $\theta_s$ value at which the pseudo-experiment was generated, here we need it for all $\theta_{s'}$ (including $s' \ne s$) and $\theta_t$.

Now we can define the critical value $\Delta \chi^2_{c,t}$ as the $w$-weighted $1-\alpha$ quantile of $\Delta \chi^2(\theta_t \mid x)$ for $x \sim \{x\}_\mathrm{mix}$, for example
\begin{equation}
	\sum_{x \in \{x\}_\mathrm{mix}} w(x \mid \theta_t) I\bigl(\Delta \chi^2(\theta_t \mid x) \ge \Delta \chi^2_{c,t}\bigr) \lesssim \alpha \times S n_\mathrm{exp}
	\label{eq:mixture:def:QF}
	,
\end{equation}
where the $\lesssim$ is meant to represent that we take the smallest $\Delta \chi^2_{c,t}$ that satisfies the inequality.

\subsection{Derivation}

In order to obtain more pseudo-experiments at large $\Delta \chi^2$ values, which would yield more precise high-CL critical values, we use an importance sampling approach: instead of directly sampling from the target distribution $p(x \mid \theta_t)$, we sample from a different distribution and weight the sampled the toys by the ratio of probability distributions to calculate the relevant quantities under the target distribution (the critical values). The question therefore becomes: what is the ideal sampling distribution to generate the desired pseudo-experiments?

Note that it is important to find a sampling distribution that is as close as possible to the target distribution apart from generating high $\Delta \chi^2$ pseudo-experiments with higher probability. In particular, if each experiment $x$ consists of $m$ measurements, the experiments are points in an $m$-dimensional space and there are $m$ dimensions in which we can stretch or shrink the sampling distribution. Instead of thinking about estimating quantiles, let's think of estimating the probability density $p(Y(x) \mid \theta_t)$ using histograms for $Y(x) \coloneq \Delta\chi^2(\theta_t \mid x)$. When using reweighting, in addition to the binomial error $n_\mathrm{exp} p(1-p)$ for the number of pseudo-experiments falling into a bin, there will be an additional contribution due to the variance of weights:
given the estimator
\begin{equation}
	\hat P_b \coloneq \frac{1}{n_\mathrm{exp}} \sum_{i=1}^{n_\mathrm{exp}} w(x_i) I(y_b \le Y(x_i) < y_{b+1})
\end{equation}
and using $I(\cdot)^2 = I(\cdot)$ we get
\begin{align}
	\mathbb E[\hat P_b]
	&= \pi_b \mathbb{E}_b[w]
	\\
	\Var[\hat P_b]
	&= \frac{1}{n_\mathrm{exp}} \left( \pi_b(1-\pi_b)\mathbb{E}_b[w]^2 + \pi_b \Var_b[w] \right)
	\label{eq:wgthist:var}
	\\
	\pi_b
	&\coloneq \mathbb{E}[I(y_b \le Y(x) < y_{b+1})]
	\\
	\mathbb{E}_b[w^k]
	&\coloneq \frac{1}{\pi_b} \mathbb{E}[w(x)^k I(y_b \le Y(x) < y_{b+1})]
	\\
	\Var_b[w]
	&\coloneq \mathbb{E}_b[w^2] - \mathbb{E}_b[w]^2
\end{align}
where in Eq.~\eqref{eq:wgthist:var} the first term is the usual binomial error due to the number of pseudo-experiments falling into the bin, and the second is the additional term due to the variance of weights among pseudo-experiments falling into the bin.

We therefore want to increase the number ($n_\mathrm{exp}\pi_b$) of pseudo-experiments falling into a high-$y(x)$ bin to reduce the binomial error, while at the same time keeping the weight-variance within the bin as small as possible. This means the ideal case of $0$-variance would be for the weights to depend on $x$ through $y(x)$ alone. Or equivalently, since the weights are the ratio of the target and sampling distribution, we want to use a sampling distribution that differs from the target distribution only by a functional factor of $\Delta \chi^2(\theta_t \mid x)$.

The key idea is to think about the meaning of a high $\Delta \chi^2(\theta_t \mid x)$ value.
The likelihood $L(\theta \mid x)$ is the probability to sample the given pseudo-experiment $x$ from $\theta$.
A high $\Delta \chi^2(\theta_t \mid x) = - 2\log L(\theta_t \mid x) / L(\hat \theta(x) \mid x)$ means there exists a value $\hat \theta(x)$ where it's more likely to sample the given pseudo-experiment than at the ``target'' $\theta_t$ value. Thus by using pseudo-experiments generated at $\theta \ne \theta_t$, we can more efficiently obtain ones with high $\Delta \chi^2(\theta_t \mid x)$.

The naive choice of simply using pseudo-experiments generated at some $\theta'$ ($\ne \theta_t$) weighted by the ratio of sampling probabilities $p(x \mid \theta_t) / p(x \mid \theta')$
however will do worse than before. This is because $\hat\theta(x)$ depends on the pseudo-experiment $x$, such that for some pseudo-experiments it may be more preferable to sample $x$ from $p(x \mid \theta_t)$ than from $p(x \mid \theta')$, resulting in an exponentially large (often unbounded) variance of weights.

The solution is simple: by using a mixture distribution $p_\mathrm{sample}(x) = \frac{1}{S} \sum_{s = 1}^S p(x \mid \theta_s)$ over a set $\{\theta\}_S \coloneq \{\theta_1, \theta_2, \cdots, \theta_S\}$ which includes $\theta_t$, we can guarantee the weights to be bounded from above, because $p_\mathrm{sample}(x) \ge \frac{1}{S} p(x \mid \theta_t)$ and hence 
\begin{equation}
	w(x \mid \theta_t) \le S
	\label{eq:mixture:deriv:weight:naivebounds}
	.
\end{equation}

\subsection{Bounds on pseudo-experiment weights for a good grid}

If we choose the grid $\{\theta\}_S$ dense \emph{and} wide enough (a \emph{good} grid) such that we may assume to have a good minimum $\hat\theta_S(x)$ on $\{\theta\}_S$ in the sense of 
\begin{align}
	\Delta\chi^2(\hat\theta_S(x) \mid x) &\le
	\begin{cases}
		\epsilon & \text{if\;} \Delta\chi^2(\theta_t \mid x) \le \Delta\chi^2_\mmax \\
		\Delta \chi^2(\theta_t \mid x) & \text{otherwise}
	\end{cases}
	\label{eq:mixture:deriv:hatcond}
	\\
	\Delta\chi^2(\hat\theta_S(x) \mid x) &\coloneq \min_s\Delta\chi^2(\theta_s \mid x)
\end{align}
for all $x$, we can put a much stricter bound on the weights than Eq.~\eqref{eq:mixture:deriv:weight:naivebounds}.
Here, $\epsilon \lesssim 1$ will be smaller for denser spacing of $\{\theta\}_S$
and $\Delta\chi^2_\mmax$ will be larger for a wider range covered by $\{\theta\}_S$.
Since this additional condition can deal with the case of $\theta_t \notin \{\theta\}_S$ as well,
let us define a symbol $C$ which is $1$ if $\theta_t \in \{\theta\}_S$ and $0$ otherwise.
Note that to guarantee Eq.~\eqref{eq:mixture:deriv:hatcond} under $C = 0$ one generally needs to have parameter values in $\{\theta\}_S$ that surround $\theta_t$ sufficiently well. For example, with a 1-dimensional continuous $\theta$ parameter, one needs $\min_s \theta_s \le \theta_t \le \max_s \theta_s$.

First, we focus on the pseudo-experiments with $\Delta\chi^2(\theta_t \mid x) \le \Delta\chi^2_\mmax$, which are our primary interest, and note that
\begin{equation}
	\frac{p(x \mid \hat\theta_S(x))}{p(x \mid \theta_t)}
	=
	\exp\big[\tfrac{1}{2} \{\Delta \chi^2(\theta_t \mid x) - \Delta \chi^2(\hat\theta_S(x) \mid x)\} \big]
	\ge 
	\exp\big[\tfrac{1}{2} \Delta \chi^2(\theta_t \mid x) - \tfrac{\epsilon}{2} \big]
	.
\end{equation}
The sum of probability ratios is now bounded from below by
\begin{equation}
	\sum_{s=1}^S \frac{p(x \mid \theta_s)}{p(x \mid \theta_t)} \ge C \times \frac{p(x \mid \theta_t)}{p(x \mid \theta_t)} + \frac{p(x \mid \hat\theta_S(x))}{p(x \mid \theta_t)} \ge C + \exp\big[\tfrac{1}{2} \Delta \chi^2(\theta_t \mid x) - \tfrac{\epsilon}{2} \big]
	\label{eq:mixture:deriv:sumprob:lowerbound}
\end{equation}
because $\{\theta\}_S$ includes both $\hat\theta_S(x)$ and (if $C = 1$) $\theta_t$.
This means for any pseudo-experiment with $\epsilon < \Delta \chi^2(\theta_t \mid x) \le \Delta\chi^2_\mmax$, it is more likely to be sampled in $S n_\mathrm{exp}$ samples from $\frac{1}{S} \sum_{s=1}^S p(x \mid \theta_s)$ than in $n_\mathrm{exp}$ samples from the target distribution $p(x \mid \theta_t)$.
The sum of probability ratios is further bounded from above by
\begin{equation}
	\sum_{s=1}^S \frac{p(x \mid \theta_s)}{p(x \mid \theta_t)} = C + \sum_{s (\ne t)} \frac{L(\theta_s \mid x)}{L(\theta_t \mid x)} \le C + (S-C) \exp\big[\tfrac{1}{2} \Delta \chi^2(\theta_t \mid x) \big]
\end{equation}
because $L(\theta_s \mid x) \le L(\hat\theta(x) \mid x)$.
This means the weights are bounded by 
\begin{equation}
	\frac{S}{C + (S-C) \exp\big[\tfrac{1}{2} \Delta \chi^2(\theta_t \mid x) \big]} \le w(x \mid \theta_t)\le \frac{S}{C + \exp\big[\tfrac{1}{2} \Delta \chi^2(\theta_t \mid x) - \tfrac{\epsilon}{2} \big]}
	\label{eq:mixture:deriv:weight:bounds}
	.
\end{equation}
We see that the bounds depend on the pseudo-experiments through $\Delta \chi^2(\theta_t \mid x)$ only, and also note that for sufficiently large $\Delta \chi^2(\theta_t \mid x)$ the ratio
of upper $w_\mmax$ to lower bound $w_\mmin$ converges to
\begin{equation}
	\frac{w_\mmax}{w_\mmin} \to (S-C) e^{\epsilon/2}
	,
\end{equation}
which indicates a small relative variance of weights as long as the number of grid points $S$ is not a very large number.

For pseudo-experiments with $\Delta\chi^2(\theta_t \mid x)$ above the threshold $\Delta \chi^2_\mmax$,
we have
\begin{equation}
	\frac{p(x \mid \hat\theta_S(x))}{p(x \mid \theta_t)}
	=
	\exp\big[\tfrac{1}{2} \{\Delta \chi^2(\theta_t \mid x) - \Delta \chi^2(\hat\theta_S(x) \mid x)\} \big]
	\ge 
	1
\end{equation}
 by Eq.~\eqref{eq:mixture:deriv:hatcond}, 
and hence an upper bound on the weights
\begin{equation}
	w(x \mid \theta_t)
	\le
	\frac{S \times p(x \mid \theta_t)}{C \times p(x \mid \theta_t) + p(x \mid \hat\theta_S(x))}
	\le
	\frac{S}{C + 1}
	\label{eq:mixture:deriv:weight:bounds:abovemax}
	.
\end{equation}

\subsection{Critical value estimator performance with a good grid}

Since quantiles (the critical values) are just the inverse function of the cumulative distribution function (CDF), 
we can estimate the relative reduction of the quantile estimation variation by the reduction of the CDF estimation variance. The relationship for an observable $y \sim f(y)$ is given by $\Var[\hat y(P)] = f(y)^{-2} \Var[\hat P(y)]$ where $\hat y(P)$ is the quantile function estimator, $\hat P(y)$ the CDF estimator, and $f(y)$ the probability distribution function.

Following Eq.~\eqref{eq:mixture:def:QF}, and using the shorthand notation $Y(x) \coloneq \Delta\chi^2(\theta_t \mid x)$, our CDF estimator is 
\begin{equation}
	\hat P(y \mid \theta_t)
	= \frac{1}{S n_\mathrm{exp}} \sum_{x \in \{x\}_\mathrm{mix}} w(x \mid \theta_t) I(Y(x) \ge y)
\end{equation}
with $x \sim \frac{1}{S} \sum_{s=1}^S p(x \mid \theta_s)$. This is an unbiased estimator for the target CDF $P(y \mid \theta_t)$
\begin{align}
	\mathbb E\big[\hat P(y \mid \theta_t)\big]
	&= \mathbb E\big[w(x \mid \theta_t) I(Y(x) \ge y)\big] \\
	&= \EE\bigl[ I(Y(x) \ge y) \bigm| \theta_t \bigr] \\
	&= P(y \mid \theta_t)
	.
\end{align}
where $\EE[\,\cdot\,]$ means to take the expectation with $x \sim \frac{1}{S} \sum_{s=1}^S p(x \mid \theta_s)$, and $\EE[\;\cdot \mid \theta_t]$ to take the expectation with $x \sim p(x \mid \theta_t)$.
Now defining 
\begin{equation}
  y_\mmax \coloneq \max\{ \Delta\chi^2_\mmax, y \},
\end{equation}
the variance from a single pseudo-experiment is
\begin{align}
	&\Var\big[w(x \mid \theta_t) I(Y(x) \ge y)\big]
	\\
	&= \mathbb{E}\big[w(x \mid \theta_t)^2 I(Y(x) \ge y)^2\big]
	 - \mathbb{E}\big[w(x \mid \theta_t) I(Y(x) \ge y)\big]^2
	\\
	&= \mathbb{E}\big[w(x \mid \theta_t) I(Y(x) \ge y) \mid \theta_t \big]
	 - P(y \mid \theta_t)^2
	\\
	&\le \mathbb{E}\biggl[\frac{S \times I(y \le Y(x) \le y_\mmax)}{C + \exp\big[\tfrac{1}{2} \big(Y(x) - \epsilon\big) \big]} \biggm| \theta_t \biggr]
	 + \frac{S}{C + 1} \mathbb{E}\bigl[I(Y(x) \ge y_\mmax) \bigm| \theta_t \bigr]
	 - P(y \mid \theta_t)^2
	\\
	&\le \frac{S}{C + \exp\big[\tfrac{1}{2} (y - \epsilon) \big]} \mathbb{E}\left[ I(y \le Y(x) \le y_\mmax) \mid \theta_t \right]
	+ \frac{S}{C + 1} P(y_\mmax \mid \theta_t)
	 - P(y \mid \theta_t)^2
	\\
	&= S \times \left[ 
	\frac{P(y \mid \theta_t) - P(y_\mmax \mid \theta_t)}{C + \exp\big[\tfrac{1}{2} (y - \epsilon) \big]} 
	+ \frac{P(y_\mmax \mid \theta_t)}{C + 1}
	\right]
	- P(y \mid \theta_t)^2
\end{align}
where in going from the second to the third line we used $I(\cdot)^2 = I(\cdot)$, and in going to the fourth line we used the upper bound from Eq.~\eqref{eq:mixture:deriv:weight:bounds} and Eq.~\eqref{eq:mixture:deriv:weight:bounds:abovemax}, and in going to the fifth line we used $Y(x) \ge y$ from the argument of the indicator function.
The variance of the CDF estimator is therefore
\begin{align}
	\Var\big[\hat P(y \mid \theta_t)\big]
	&= \frac{1}{S n_\mathrm{exp}} \Var\big[w(x \mid \theta_t) I(\Delta\chi^2(\theta_t \mid x) \ge y)\big]
	\\
	&\le \frac{1}{n_\mathrm{exp}} \left(
	  \frac{P(y \mid \theta_t) - P(y_\mmax \mid \theta_t)}{C + \exp\big[\tfrac{1}{2} (y - \epsilon) \big]} 
		+ \frac{P(y_\mmax \mid \theta_t)}{C + 1} 
		- \frac{P(y \mid \theta_t)^2}{S}
	\right)
\end{align}
where we note that the $S$ factors in the first two terms were cancelled thanks to being able to reuse the pseudo-experiments generated at all $S$ values for the CDF estimation of each $\theta_t$ value.

\begin{figure}[tp]
	\centering
	\subfloat[Upper bound on $\gamma$]{\label{fig:mixture:perf:gamma}%
	  \includegraphics[width=0.49\textwidth]{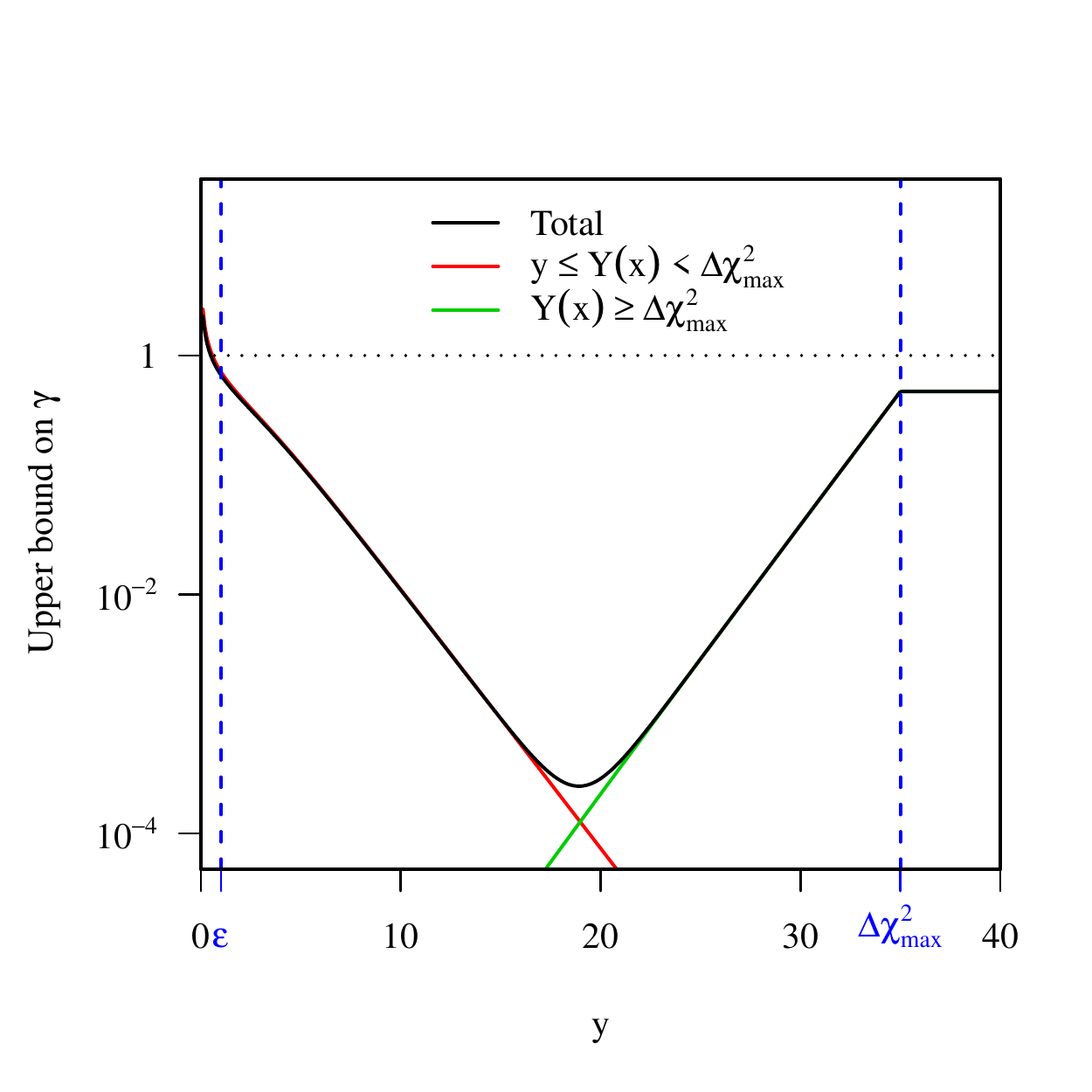}}%
	\subfloat[Relative error on estimated CDF]{\label{fig:mixture:perf:CDFerr}%
	  \includegraphics[width=0.49\textwidth]{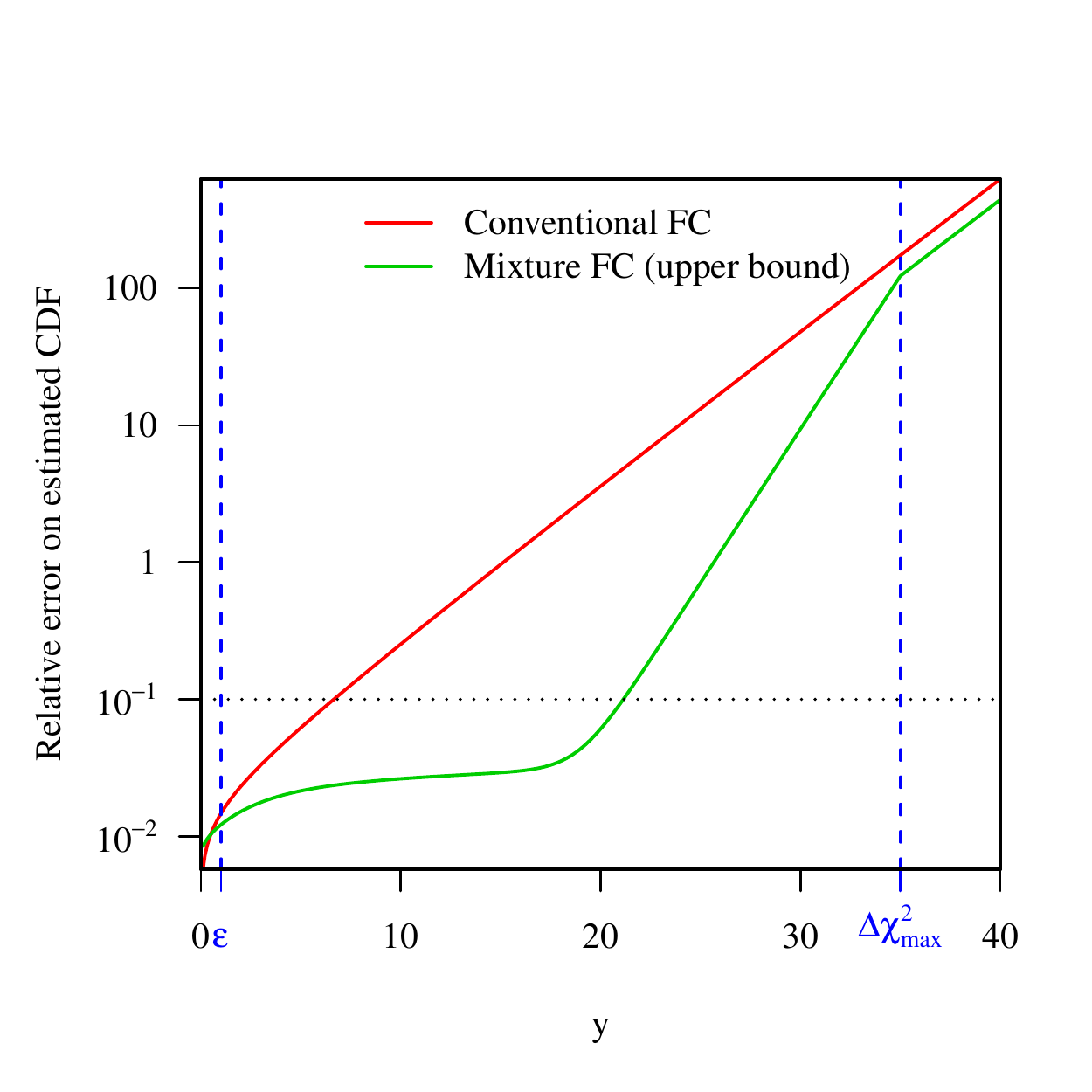}}%
	\caption{
	(a) Example functional shape of upper bound on the ratio of estimated critical value variance from Eq.~\eqref{eq:mixture:perf:gamma}. The red line indicates the error contribution from pseudo-experiments with $y \le Y(x) < \Delta\chi^2_\mmax$ (first term with $A(y)$) which is responsible for the exponential reduction of total uncertainty until the contribution from pseudo-experiments with $y \ge \Delta\chi^2_\mmax$ (second term with $B(y)$ shown by green line) takes over for very high CL critical values.
	(b) The relative error on the calculated CDF estimator $\hat P(y \mid \theta_t)$ assuming $n_\mathrm{exp} = 10,000$ pseudo-experiments at each sampling value $\theta_s$. A reference 10\% error threshold is indicated by the dotted line.
	The example used for both plots is constructed assuming $\epsilon = 1$, $S = 10$, $C=1$, $\Delta\chi^2_\mmax = 35$ and the true $Y(x)$-CDF is assumed to be chi-squared with 1~degree of freedom. In (a), the exponential growth factor for the green line depends on the assumed CDF, unlike the red line whose decay factor is given by Eq.~\eqref{eq:mixture:perf:gamma}.}
	\label{fig:mixture:perf:gammaCDFerr}
\end{figure}

For reference, the variance on the CDF estimator in the conventional FC method (denoted in the following equations by ``conv'') is given by the binomial error
\begin{equation}
	\Var[\hat P_\mathrm{conv}(y) \mid \theta_t] = \frac{1}{n_\mathrm{exp}} \left(
	P(y \mid \theta_t) - P(y \mid \theta_t)^2
\right)
,
\end{equation}
so the variance on the estimated critical values $\hat y(P \mid \theta_t)$ in the mixture-FC method is smaller by the factor
\begin{align}
	\gamma
	&\coloneq 
	\frac{
   	\Var[\hat y(P \mid \theta_t)]
	}{
	  \Var[\hat y_\mathrm{conv}(P \mid \theta_t) \mid \theta_t]
	}
	\\
	&=
	\frac{
   	\Var\big[\hat P(y \mid \theta_t)\big]
	}{
	  \Var\big[\hat P_\mathrm{conv}(y \mid \theta_t) \mid \theta_t\big]
	}
	\\
	&\le \frac{
	  A(y) + B(y) P(y_\mmax \mid \theta_t) / P - \frac{1}{S} P
	}{
	  1 - P
	}
	\label{eq:mixture:perf:gamma}
	\\
	A(y) &\coloneq \frac{1}{C + \exp\big[\tfrac{1}{2}(y - \epsilon) \big]} \\
	B(y) &\coloneq \frac{1}{C+1} - A(y)
	.
\end{align}
where $y$ is the true $P$-quantile satisfying $P(y \mid \theta_t) = P$.
The typical functional shape of the upper bound is shown in Fig.~\ref{fig:mixture:perf:gamma}.
Let us first consider the case of $P \ll P(y_\mmax \mid \theta_t)$.
For the small $P \le 1/2$ values one is typically interested in, 
the mixture model method obtains more precise critical values than the conventional method (i.e.~$\gamma \le 1$)
for all $y \ge \epsilon$ if $\theta_t \in \{\theta\}_S$ ($C=1$), or all $y \ge \epsilon + 2 \log 2$ if $\theta_t \notin \{\theta\}_S$ ($C=0$).
As $y$ increases, the relative variance first decreases linearly, and for $y \gtrsim 2 + \epsilon$ it starts to decrease exponentially as $\gamma \lesssim \exp(-y/2)$.
As $y$ further increases toward $\Delta \chi^2_\mmax$, and $P \ll P(y_\mmax \mid \theta_t)$ fails to hold anymore, the $B(y) P(y_\mmax \mid \theta_t)/P$ term becomes dominant,
which saturates to $P(y_\mmax \mid \theta_t)/P = 1$ for $y \ge \Delta \chi^2_\mmax$.
Hence the improvement flattens out to $\gamma \le \frac{1}{C+1}$ for $y \ge \Delta \chi^2_\mmax$, which is still at least as good as the conventional FC method. By choosing suitable parameter points $\{\theta\}_S$ and thus a suitable $\Delta \chi^2_\mmax$, critical values of the desired precision can be calculated.
As the exponential reduction in variance cancels the typically exponential dependence of the CDF on the test-statistic ($\exp(-y/2)$ in the case of a chi-squared distribution), the relative error on the estimated CDF becomes approximately flat over a wide range of test-statistic values (Fig.~\ref{fig:mixture:perf:CDFerr}), which is much more efficient than for the conventional FC where low-CL become over-precise with more pseudo-experiments, while high-CL still suffer from large errors.

\subsection{Interpolation}

While for the conventional FC method one can only compute the critical values at the parameter value $\theta_s$ where the pseudo-experiments were generated, in the mixture-FC method it is sufficient to guarantee that the target parameter value $\theta_t$ is sufficiently close and surrounded by the sampling points $\{\theta\}_S$ such that condition Eq.~\eqref{eq:mixture:deriv:hatcond} holds. Considering that for a typical setup the toys to be generated are the same as those used in the conventional FC method, this means that the mixture-FC method not only reduces the uncertainty on the critical values at the sampling points $\{\theta\}_S$, but also allows interpolating the critical values between these points with similar performance.

\subsection{Diagnostics and error estimation}

As the mixture-FC method exploits the relationship of the $\Delta \chi^2$ statistic to the probability of sampling pseudo-experiments, it is essential that the calculation of $\Delta \chi^2$ matches the process used to generate the pseudo-experiments. It is for example not allowed to sample from a poisson random number generator while using an approximation like Pearson's $\chi^2$ for $\Delta\chi^2$. A simple diagnostic is to calculate the average weight across all pseudo-experiments in $\{x\}_\mathrm{mix}$ and check that this is equal to $1$ up to statistical fluctuations. Since the same pseudo-experiments will be used for all target parameter values $\theta_t$ (which the weights are a function of), the statistical fluctuations of these average weights will be correlated for different $\theta_t$ values.

To estimate the error of the computed critical values, we recommend using resampling methods such as the non-parametric bootstrap~\cite{10.1214/aos/1176344552} or jackknife~\cite{10.2307/2334280} instead of simple methods like binomial errors, in order to capture not only the statistical fluctuations in the number of pseudo-experiments that fall into a range of $\Delta \chi^2$ values, but also the statistical fluctuations in their weights.

\section{Example with a single cyclic parameter}

\begin{figure}[tp]
	\centering
	\includegraphics[width=0.49\textwidth]{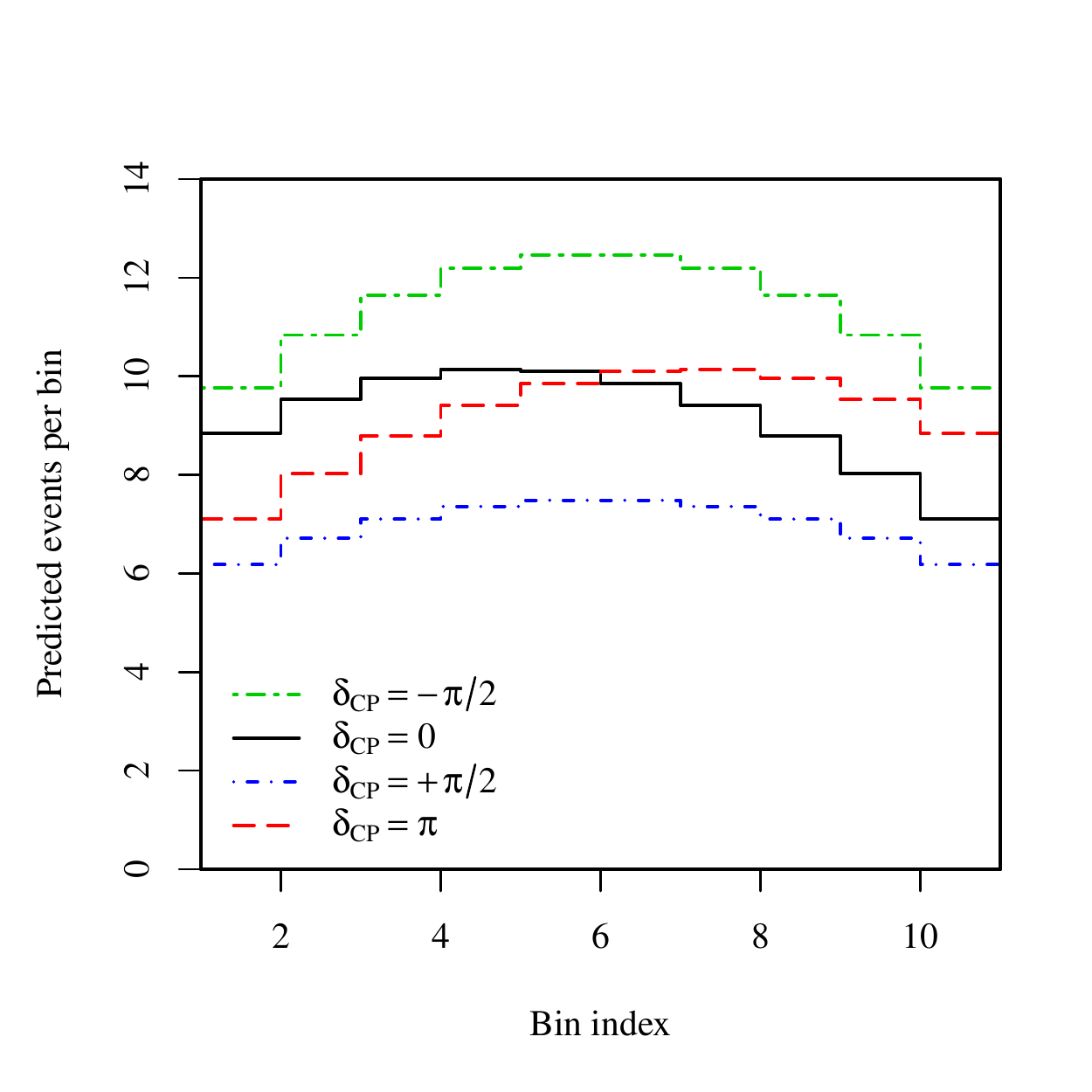}%
	\caption{Predicted number of events $\lambda_b(\delta_\mathrm{CP})$ for each bin index $b$ as used in the example.}
	\label{fig:example:CP:lambda}
\end{figure}

We consider a simple example that uses a binned-Poisson model, inspired by the search for CP violation in a long-baseline neutrino oscillation experiment, here in particular the T2K experiment~\cite{T2K:2011qtm}.
The model has a single angular parameter called the ``CP violation phase'' $\delta_\mathrm{CP} \in [-\pi,\pi]$ which is constrained by $B=10$ Poisson-distributed observations $n_b \sim \mathrm{Poisson}(\lambda_b)$ with the predicted event rate
\begin{align}
	\lambda_b(\delta_\mathrm{CP}) &\coloneq 10 \times ( 1 - \phi_b^2 ) \times \left( 1 - \frac{1}{4} \sin(\delta_\mathrm{CP} + \phi_b) \right)
	\\
	\phi_b &\coloneq \frac{b - 5.5}{10}
\end{align}
for each bin with index $b = 1, 2, \cdots, 10$ (Fig.~\ref{fig:example:CP:lambda}). The main feature of this model is that one is mostly sensitive to $\sin \delta_\mathrm{CP}$ through the overall normalization of approximately 100~total observations ($\sum_b n_b$), and weakly sensitive to the $\cos \delta_\mathrm{CP}$ component through the ``shape'' of the observations as a function of $b$ (meant to represent bins of increasing neutrino energy).
Deviations from Wilks' theorem are caused by $\sin \delta_\mathrm{CP}$ having physical boundaries at $\pm 1$ (resulting in reduced critical values around $\sin\delta_\mathrm{CP} = \pm 1$), the sign of $\cos \delta_\mathrm{CP}$ acting as an effectively discrete degree of freedom (resulting in increased critical values at some $\sin\delta_\mathrm{CP} \ne \pm 1$ values), as well as the Poisson nature of the observations. In an actual experiment, one would have further continuous discrete physics parameters degenerate with $\delta_\mathrm{CP}$ as well as various systematic uncertainties treated as nuisance parameters. For simplicity and clarity however we focus on $\delta_\mathrm{CP}$ alone, which for continuity with the earlier sections will be referred to as $\theta = (\delta_\mathrm{CP})$, and the observations as $x = (n_1, n_2, \cdots, n_{10})$.

\begin{figure}[tp]
	\centering
	\subfloat[$\Delta \chi^2_t$ at various $\theta_s$]{\label{fig:example:CP:toys:theta}%
	  \includegraphics[width=0.49\textwidth]{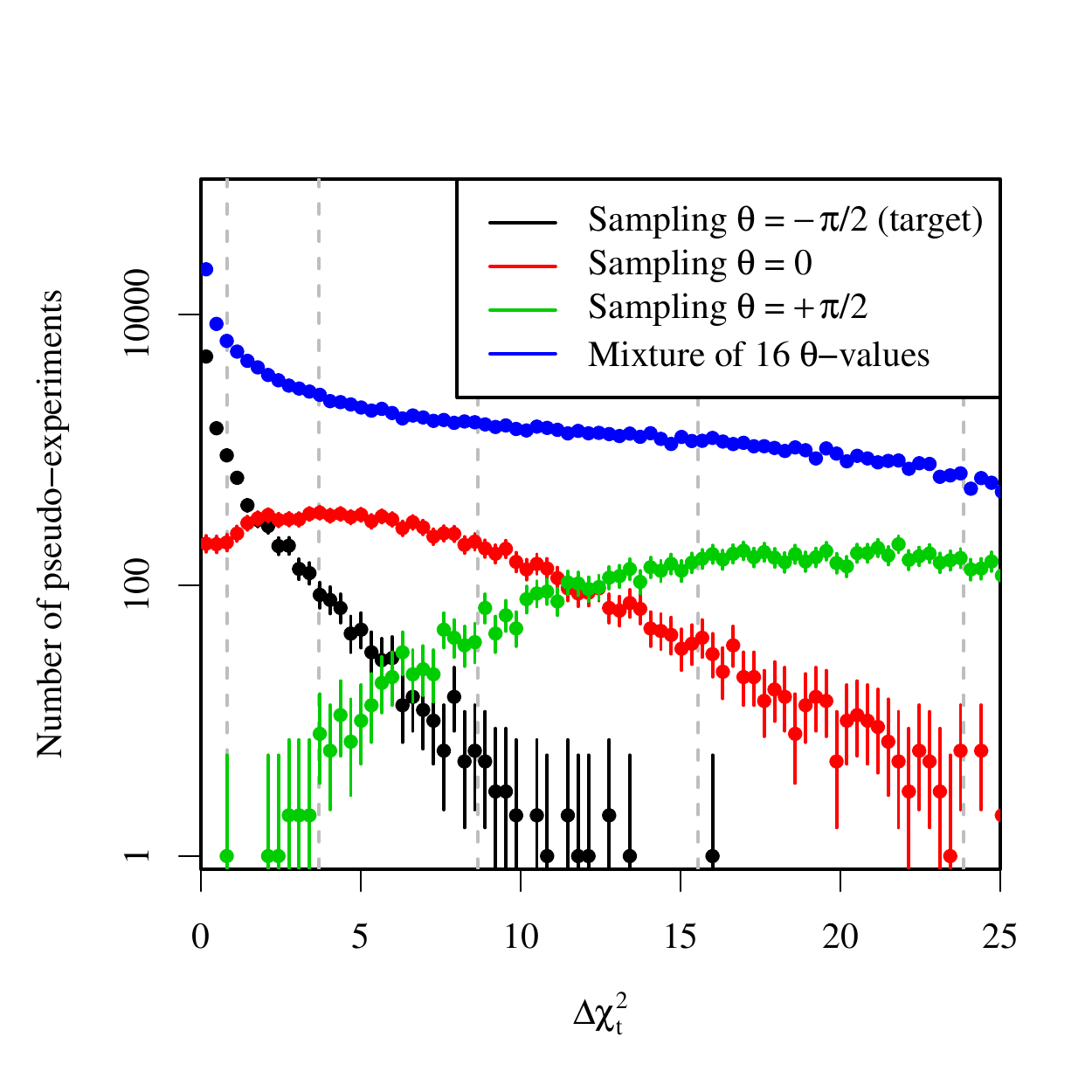}}%
		%
	\subfloat[Estimated $\Delta \chi^2_t$ distributions at $\theta_t$]{\label{fig:example:CP:toys:std-vs-mixture}%
	   \includegraphics[width=0.49\textwidth]{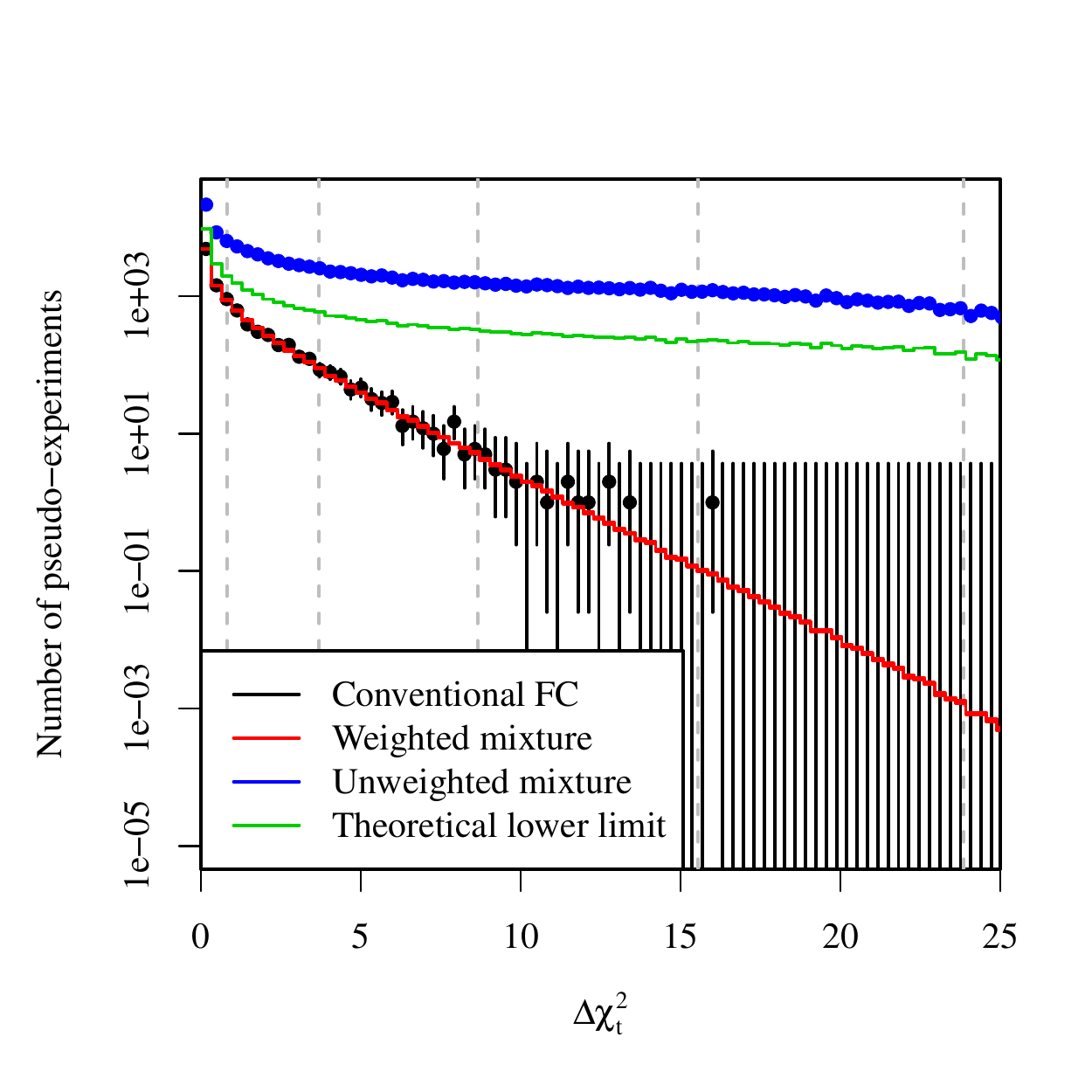}}%
	\caption{$\Delta \chi^2_t \coloneq \Delta \chi^2(\theta_t \mid x)$ distributions with target $\theta_t = -\pi/2$ for (a) various sampling parameter values $\theta_s$, and (b) comparison of estimated distributions at $\theta_t$ obtained using standard FC and mixture-FC methods. In both plots, error bars indicate $1\sigma$ binomial confidence intervals. Vertical dashed lines indicate $1,2,3,4,5\sigma$ confidence level critical values obtained by the mixture-FC method. (a) Error bars are omitted for bins with zero entries for clarity. (b) The red ``weighted mixture'' histogram is also drawn with boxes representing the error from number of pseudo-experiments in each bin and their weight variance, but these errors are smaller than the line width and not visible. The ``theoretical lower limit'' on the total number of pseudo-experiments in the mixture distribution is obtained by multiplying the lower bound in Eq.~\eqref{eq:mixture:deriv:sumprob:lowerbound} to the red ``weighted mixture'' histogram assuming $C = 1$ and $\epsilon = \epsilonValue$.}
\end{figure}

We generate $n_\mathrm{exp} = 10,000$ pseudo-experiments at each of $S=16$ values of $\theta$ evenly distributed in the parameter range $[-\pi,\pi]$.
We first focus on the target value of $\theta_t = -\pi/2$.
Fig.~\ref{fig:example:CP:toys:theta}
shows the distribution of $\Delta\chi^2(\theta_t \mid x)$ obtained for pseudo-experiments $x$ sampled from different $\theta_s$ values. In the conventional FC method, only those generated at $\theta_s = \theta_t$ are used, which correspond to the black histogram which falls off quickly for large $\Delta\chi^2(\theta_t \mid x)$.
In the mixture-FC method we further make use of the pseudo-experiments generated at all other $\theta_s$ values, of which $\theta_s = 0$ and the other extreme of $\theta_s = \pi/2$ are shown by the red and green histograms respectively. Clearly, the pseudo-experiments sampled from the shifted $\theta_s$ values have a significantly higher fraction of large $\Delta\chi^2(\theta_t \mid x)$ values.
At the same time, one can see one of the problems arising from using only the pseudo-experiments generated at $\theta_s = \pi/2$, in that one would need to apply very large weights for the small $\Delta \chi^2(\theta_t \mid x)$ region where $\theta_s = \pi/2$ has a very small sampling probability.
The mixture of pseudo-experiments generated at all 16~$\theta$ values however, shown by the blue histogram, 
is able to generate more pseudo-experiments for all $\Delta\chi^2(\theta_t \mid x)$ values, with the difference in slope compared to the black target histogram showing the exponential increase is pseudo-experiments for larger $\Delta\chi^2(\theta_t \mid x)$ values. This is even clearer to see in Fig.~\ref{fig:example:CP:toys:std-vs-mixture} where the mixture distribution was reweighted using the assigned weights. Good agreement with the target distribution as simulated by the conventional FC method is seen, and the total number of unweighted pseudo-experiments in the mixture-FC method exceeds the theoretical lower bound.

\begin{figure}[tp]
	\centering
	\subfloat[Distribution of weights]{\label{fig:example:CP:toys:weights}%
	   \includegraphics[width=0.49\textwidth]{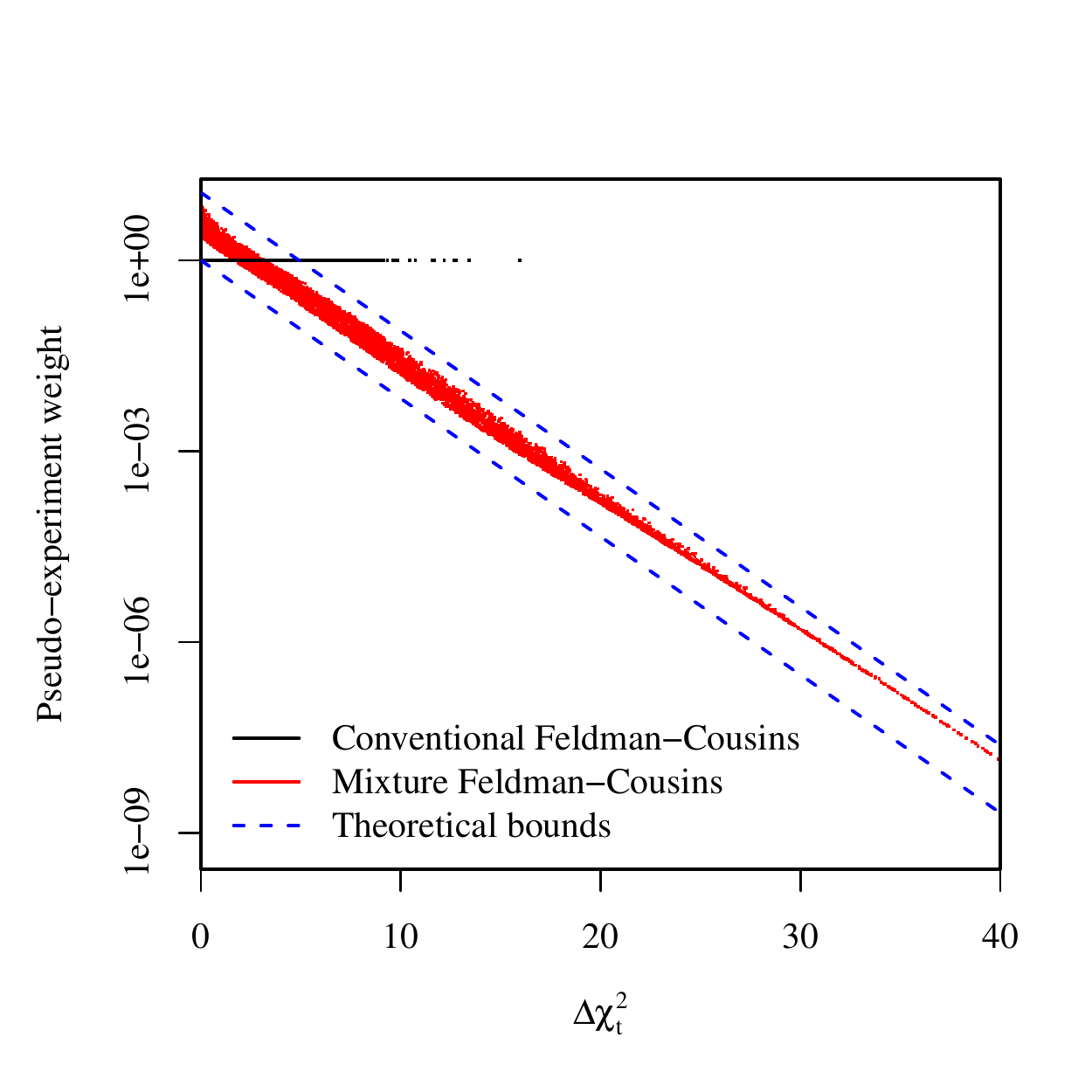}}%
	\subfloat[Distribution of $\Delta\chi^2(\hat\theta_S \mid x)$]{\label{fig:example:CP:toys:epsilon}%
	  \includegraphics[width=0.49\textwidth]{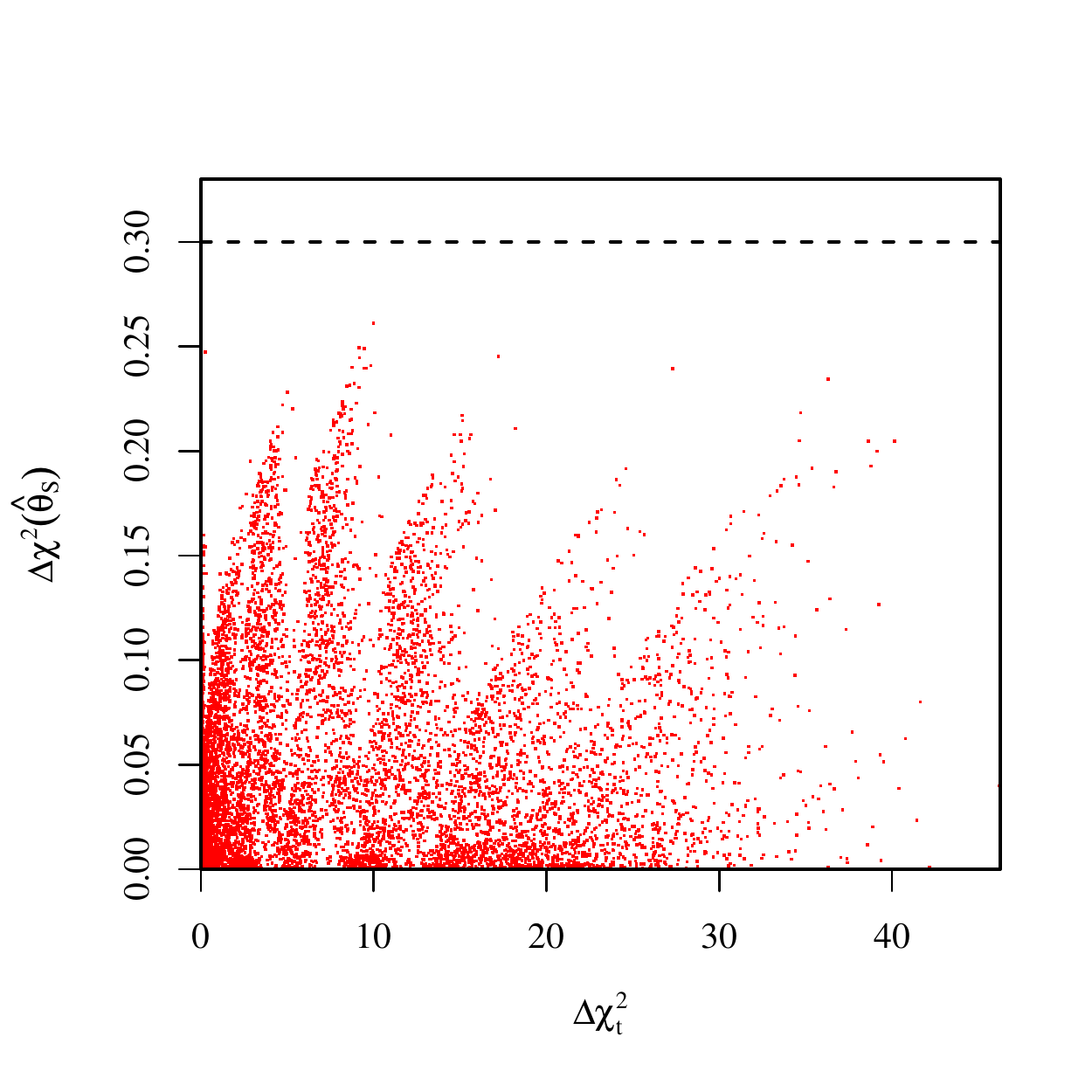}}%
	\caption{Distribution of (a) weights $w(x \mid \theta_t)$ and (b) minimum $\Delta\chi^2$ values over $\{\theta\}_S$ for each pseudo-experiment as function of $\Delta\chi^2_t \coloneq \Delta\chi^2(\theta_t\mid x)$. (a) Dashed blue lines indicate theoretical limits assuming $\epsilon = \epsilonValue$ and $C=0$. (b) The dashed line indicates $\epsilon = \epsilonValue$.}
\end{figure}

\begin{figure}[tp]
	\centering
	\includegraphics[width=0.49\textwidth]{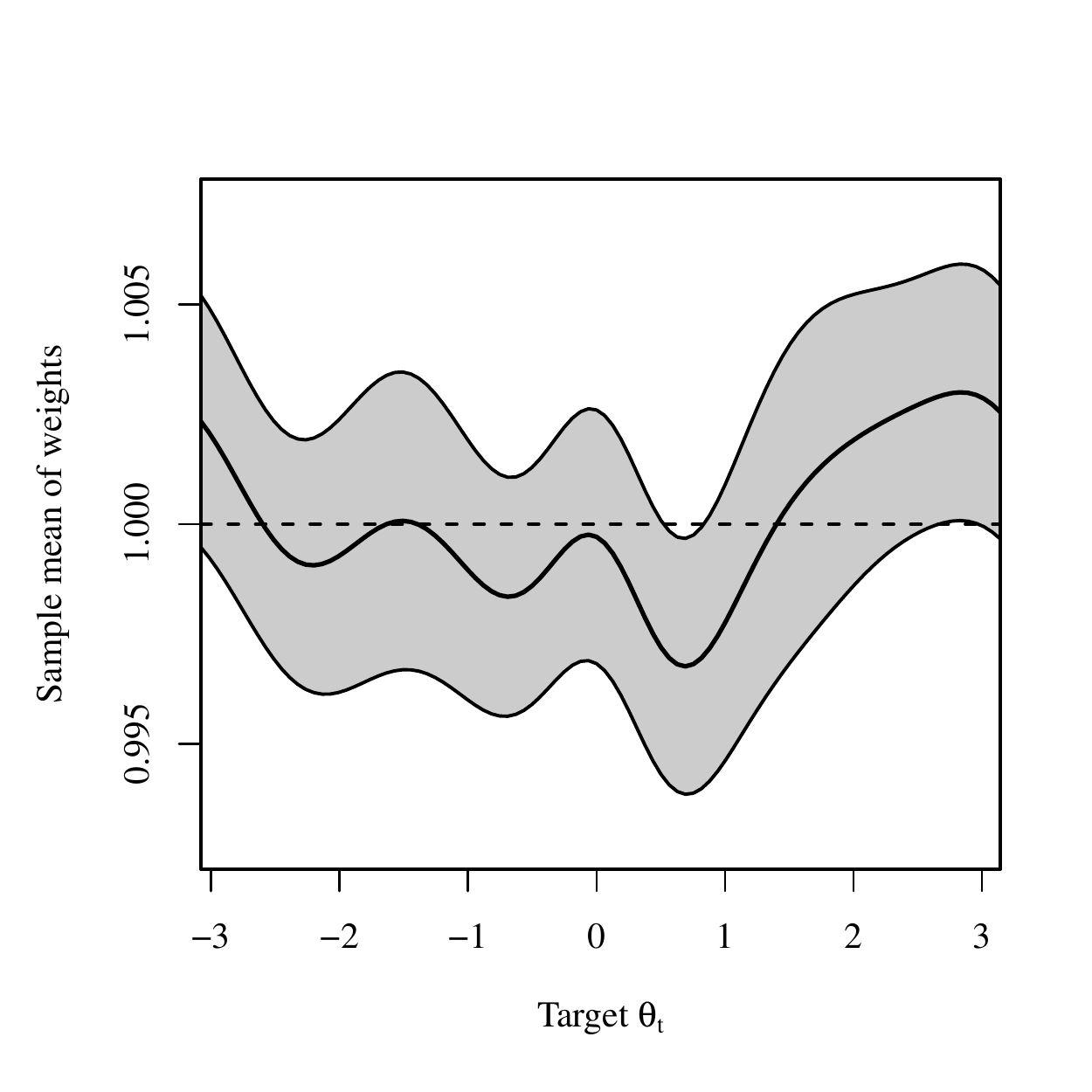}%
	\caption{The sample mean of mixture-FC pseudo-experiment weights $w(x \mid \theta_t)$ as a function of the target $\theta_t$ value. The error bands indicate the $1\sigma$ standard error on the mean, which are correlated between different target $\theta_t$ values.}
	\label{fig:example:CP:toys:sum-of-weights}
\end{figure}

\begin{figure}[tp]
	\centering
	\subfloat[$1$--$5\sigma$ CL]{\label{fig:example:CP:critical:overview}%
	  \includegraphics[width=0.49\textwidth]{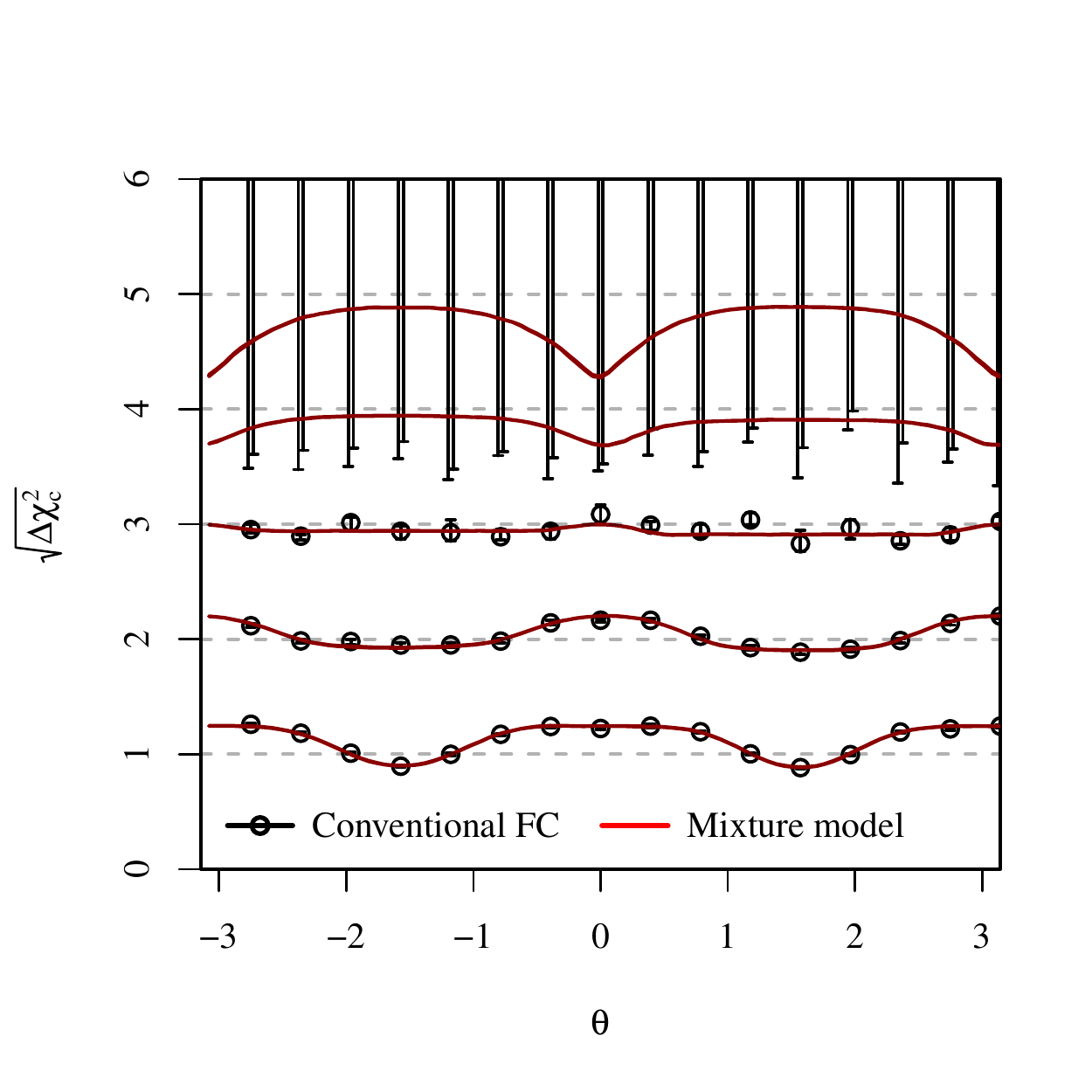}}%
	\subfloat[$1\sigma$ CL]{\label{fig:example:CP:critical:1sig}%
	   \includegraphics[width=0.49\textwidth]{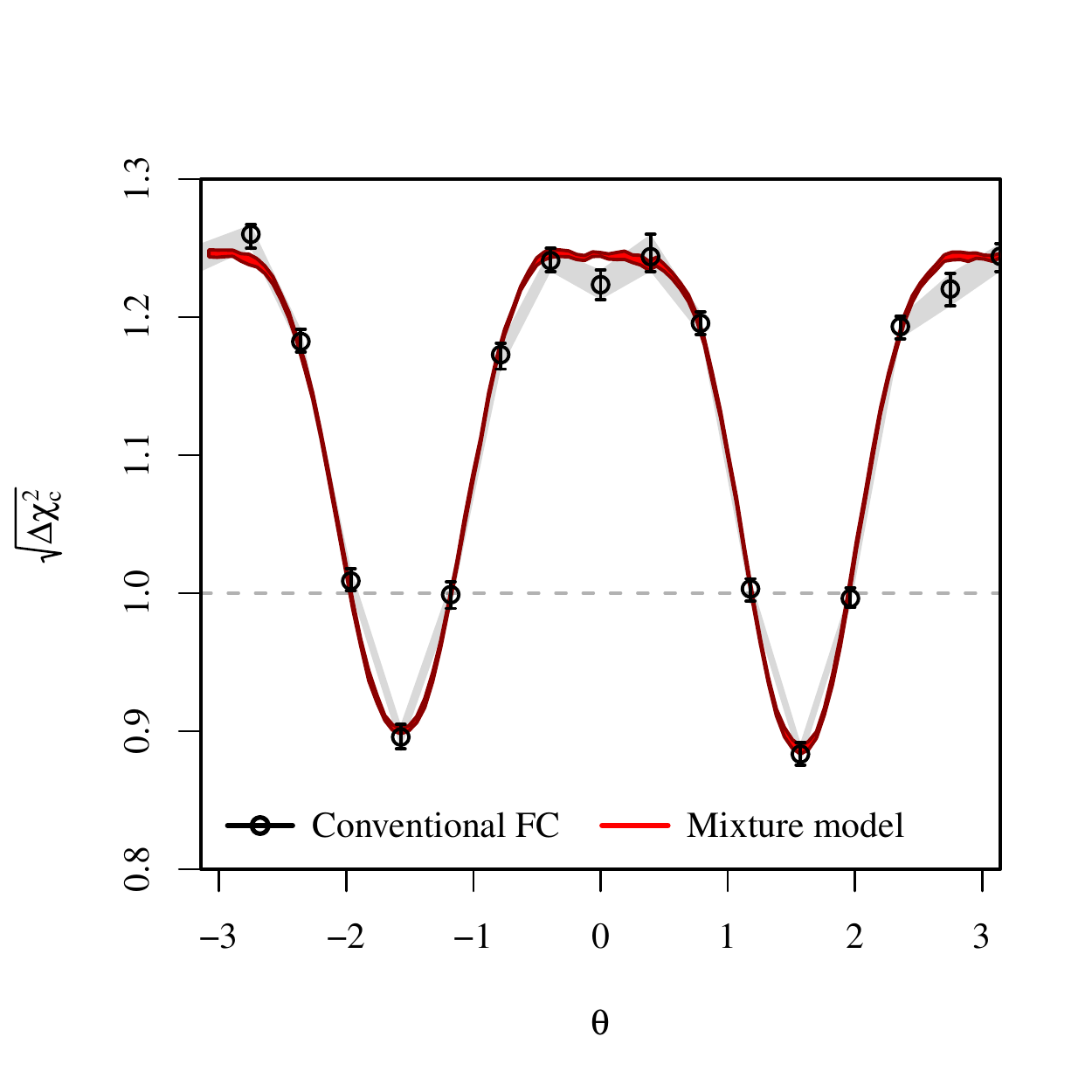}}%
		\\
	\subfloat[$2\sigma$ CL]{\label{fig:example:CP:critical:2sig}%
	   \includegraphics[width=0.49\textwidth]{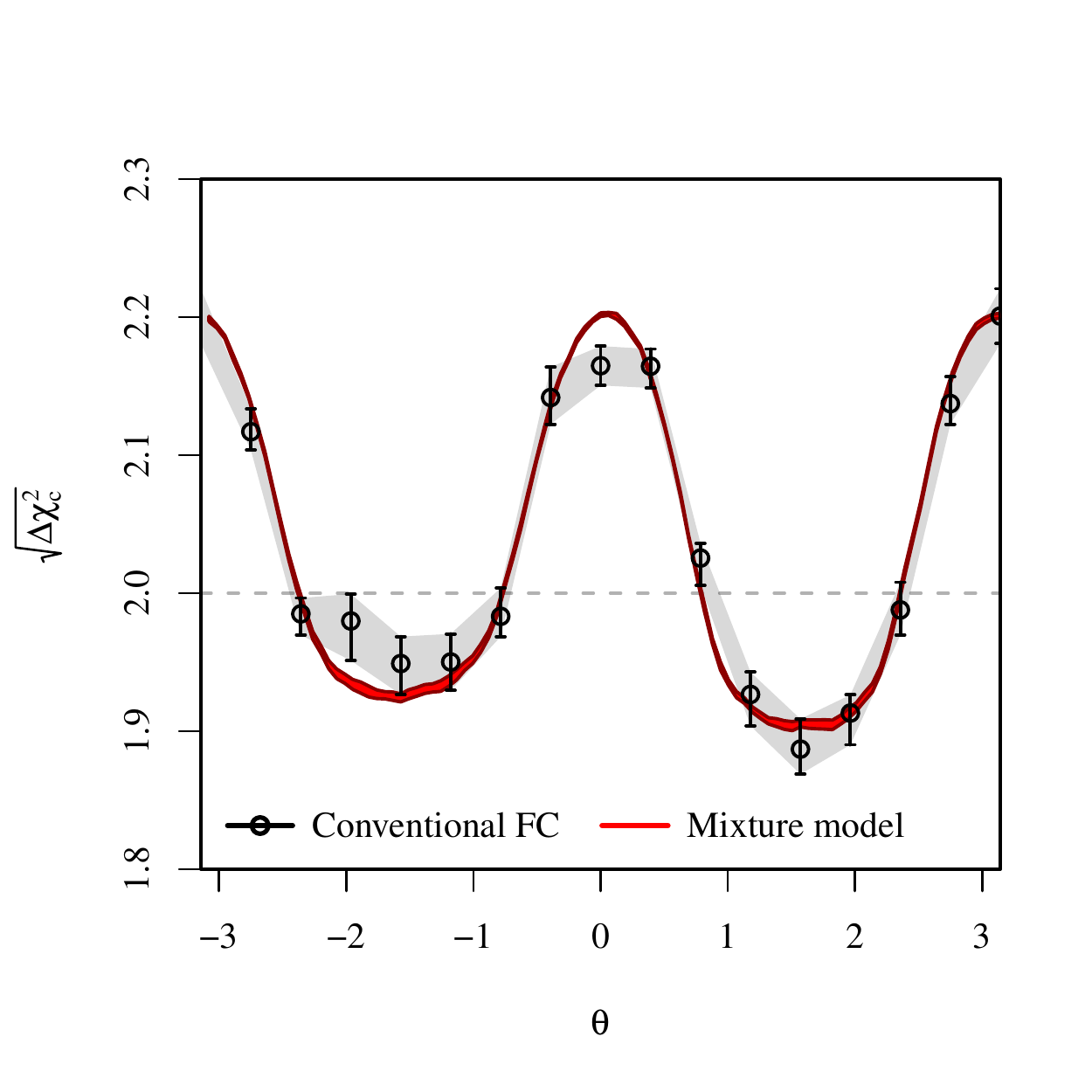}}%
	\subfloat[$3\sigma$ CL]{\label{fig:example:CP:critical:3sig}%
	   \includegraphics[width=0.49\textwidth]{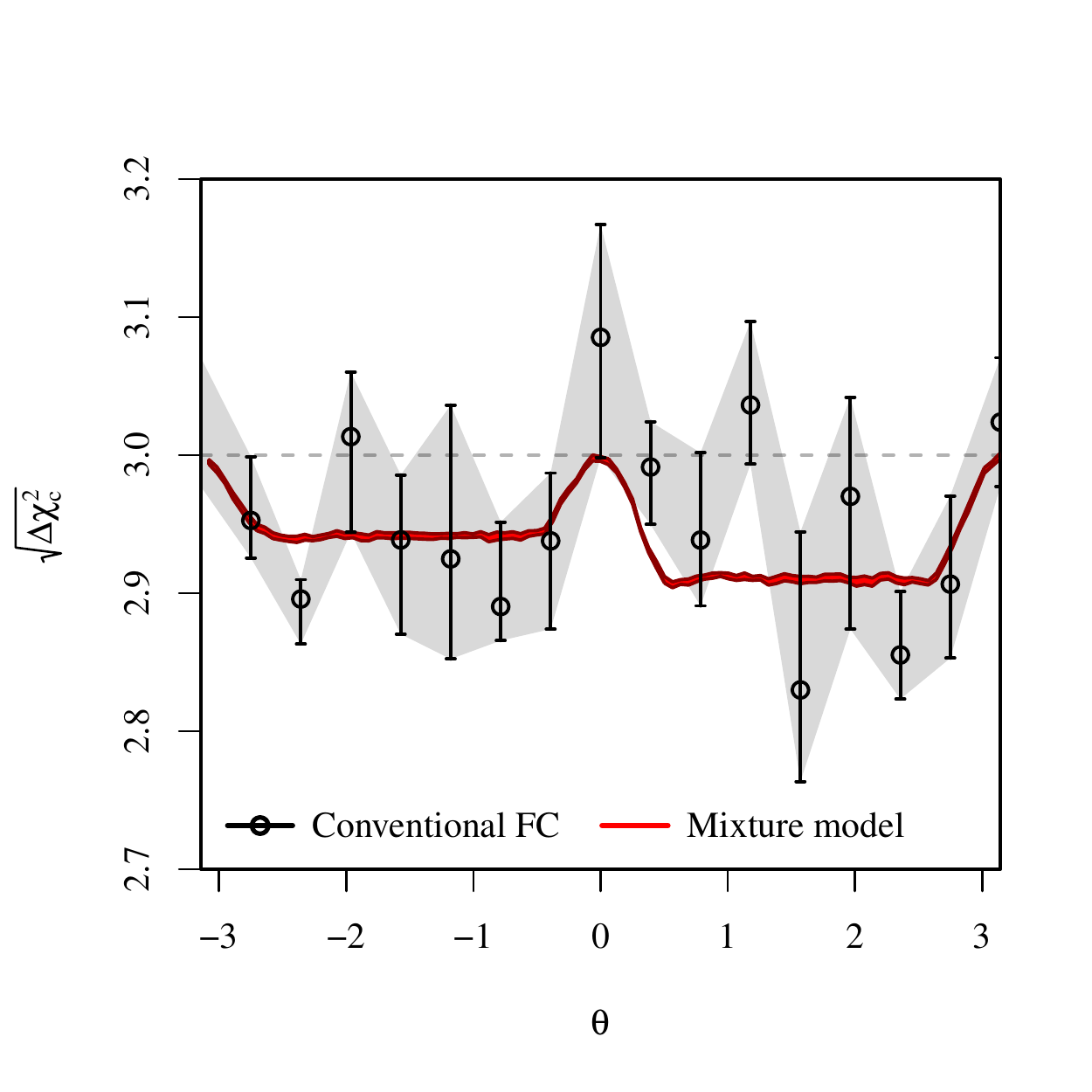}}%
	%
	\caption{$1,2,3,4,5\sigma$ confidence level critical values (given as $\sqrt{\Delta\chi^2_c}$) obtained from the same set of pseudo-experiments with the standard FC (black error bars) and mixture-FC method (red error bands). In both cases the error bars/bands indicate the $1\sigma$ error on the csritical values obtained with binomial/bootstrap errors for the standard/mixture FC method respectively. Dashed lines indicate critical values by Wilks' theorem, which are not valid here, but still drawn for reference. (a) The black error bars for $4$ and $5\sigma$ have been slightly offset to prevent overlap. (b,c,d) The gray error bands indicate the linear interpolation of the error bar end-points in the standard FC method.}
	\label{fig:example:CP:critical}
\end{figure}

\begin{figure}[tp]
	\centering
	\includegraphics[width=0.49\textwidth]{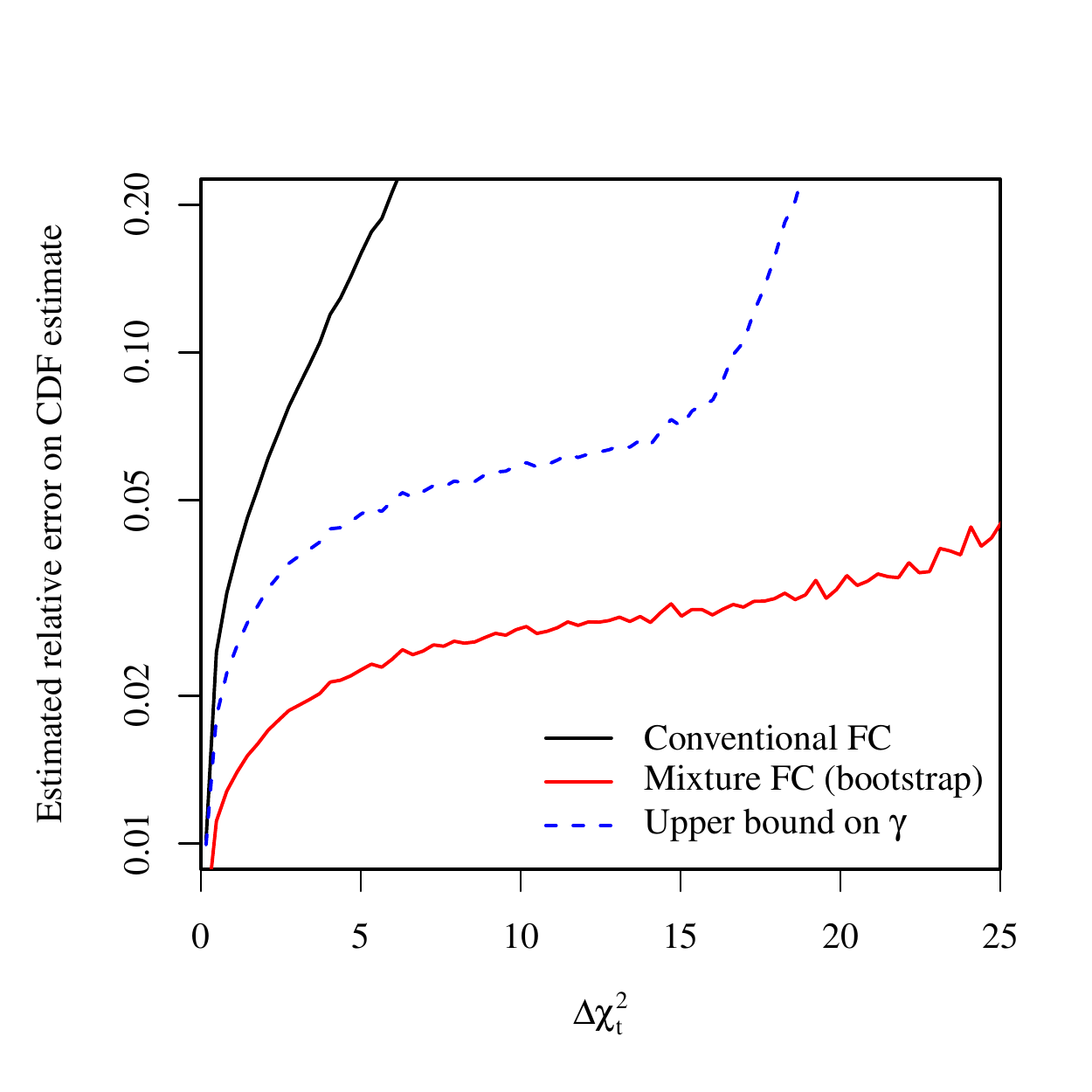}%
	\caption{Estimated relative errors on the CDF estimator $\hat P(y \mid \theta_t)$ with target $\theta_t = -\pi/2$. For the standard FC the standard error from the binomial distribution is shown (black solid line), where the more precise CDF estimate from the mixture-FC method was used in computing these errors.
	For the mixture-FC method the bootstrap error estimate (red solid line) is well below the theoretical upper limit of Eq.~\eqref{eq:mixture:perf:gamma} calculated assuming $\epsilon = \epsilonValue$ and $\Delta\chi^2_\mmax = \DchimaxValue$ (blue dashed line).}
	\label{fig:example:CP:CDFrelerr}
\end{figure}

We now check some of the diagnostics for the mixture-FC method.
The distribution of importance sampling weights $w(x \mid \theta_t)$ are shown in 
Fig.~\ref{fig:example:CP:toys:weights}
which are found to be mostly a function of $\Delta\chi^2(\theta_t \mid x)$ with small additional variance.
The weights are found to be well contained by the theoretical bounds from Eq.~\eqref{eq:mixture:deriv:weight:bounds}, which were drawn assuming a $\epsilon = \epsilonValue$ value by looking at the $\Delta \chi^2(\hat\theta_S(x) \mid x)$ distributions in Fig.~\ref{fig:example:CP:toys:epsilon}.
The sum of weights is found to be consistent with $1$ (Fig.~\ref{fig:example:CP:toys:sum-of-weights}).

Next, we look at the critical values. Fig.~\ref{fig:example:CP:critical} shows the critical values as function of the (true/target) parameter value $\theta$ using both the standard FC method (black error bars) and the mixture-FC method. Despite using the same set of pseudo-experiments, the critical values obtained with the mixture-FC method have significantly smaller uncertainty especially at higher CL, and also provide access to details of the functional shape between the 16~sampling values of $\theta$.

For the $1\sigma$ critical values (Fig.~\ref{fig:example:CP:critical:1sig}) we see that despite the relatively fine spacing of sampling values, the interpolation error as indicated by the non-overlap of red and gray error bands next to the $\theta = \pm\pi/2$ values is larger than the size of the binomial error band in the conventional method. As these binomial error bands do not capture the interpolation error, their smallness can be misleading, and renders the interpolation feature of the mixture-FC method very useful. 

For the $2\sigma$ (Fig.~\ref{fig:example:CP:critical:2sig}) and $3\sigma$ critical values (Fig.~\ref{fig:example:CP:critical:3sig}) we see good consistency between the two methods while also noting the significantly smaller errors in the mixture-FC calculation. For $3\sigma$ CL (Fig.~\ref{fig:example:CP:critical:3sig}) we see the errors in the conventional method are already so large that some of the features of the critical values are not recognizable, such as the bumps at $\theta = 0,\pi$ and the asymmetry of critical values for a flip of the $\sin\delta_\mathrm{CP}$ sign, caused by Poisson statistics.

For $4\sigma$ and higher CL the conventional FC method is unable to determine the critical values except for a lower limit. The mixture-FC method on the other hand still produces critical values with comparable relative error sizes to the lower CL critical values.

The estimated relative errors are plotted in Fig.~\ref{fig:example:CP:CDFrelerr} and are consistent with the typical shape from theoretical arguments (Fig.~\ref{fig:mixture:perf:CDFerr}). To draw the upper bound from $\gamma$ in Eq.~\eqref{eq:mixture:perf:gamma}, we conservatively assume $\Delta\chi^2_\mmax = \DchimaxValue$ based on Fig.~\ref{fig:example:CP:toys:epsilon}, i.e.\ we will only assume $\Delta \chi^2(\hat\theta_S(x) \mid x) \le \epsilon$ up to $\Delta\chi^2(\theta_t \mid x) \le \DchimaxValue$.
In this example, the actual mixture-FC error estimated with the bootstrap is smaller than the theoretical upper limit from $\gamma$ by about factor~2 for $\Delta\chi^2_t < 16$. 
This can be interpreted as more than one sampling value $\theta_s$ contributing to the sampling of each pseudo-experiment, rather than the assumption in the theoretical upper limit that only $\hat\theta_S(x)$ would contribute.
For $\Delta\chi^2_t > 16 = \Delta\chi^2_\mmax/2$ on the other hand the theoretical upper limit starts to increase significantly, whereas the actual error estimated with the bootstrap only grows slowly.
This can be interpreted as our choice of $\Delta\chi^2_\mmax = \DchimaxValue$ being overly conservative: with the present example, the chosen sampling grid $\{\theta\}_S$ appears to be effective up to significantly higher $\Delta \chi^2$ values. This is partly due to the convenient situation of having a parameter $\theta = (\delta_\mathrm{CP})$ with a bounded parameter space $\delta_\mathrm{CP} \in [-\pi,\pi]$.

\section{Discussion}

\subsection{Relation to techniques in statistical mechanics}

The presented method is similar in spirit to the ``Multiple Histogram Reweighting'' (``multi-histogram'') method~\cite{Ferrenberg:1989ui} in statistical mechanics, where statistical ensembles are simulated for various parameter values and combined by reweighting to the desired parameter value. In the multi-histogram method, the ensembles are combined with an additional per-ensemble weight, which is adjusted to minimize the overall error on the variable to be estimated.
A similar per-ensemble weighting could be applied in the presented mixture-FC method as well, where these additional weights would be allowed to depend on the target $\Delta \chi^2_t \coloneq \Delta \chi^2(\theta_t \mid x)$ value as well, in order to reduce the variance on the critical value estimator as much as possible.

One difference to the multi-histogram method however, is that because we do not resort to Markov-Chain Monte-Carlo techniques to sample the pseudo-experiments, the sampling distribution of pseudo-experiments $x$ at each parameter value $\theta$ is known exactly including the normalization constant. Hence the iterative procedure that is required at the end of the multi-histogram method to self-consistently determine these normalization constants (the free energies) is not necessary in the mixture-FC method.

 \subsection{Relation to the marginal distribution}

The sampling distribution constructed as a mixture over several parameter values $\{\theta\}_S$ can be considered a marginal probability distribution with prior $\pi(\theta) = \frac{1}{S} \sum_{s=1}^S \delta(\theta - \theta_s)$, where $\delta(\cdot)$ is the Dirac delta function.
Additional per-ensemble weights as discussed in the previous paragraph would correspond to
an alternative prior $\pi(\theta \mid \Delta\chi^2_t) = \sum_{s=1}^S r_s(\Delta\chi^2_t) \, \delta(\theta - \theta_s)$ where $r_s(\Delta\chi^2_t)$ can be optimized to reduce errors subject to the condition $\sum_s r_s(\Delta\chi^2_t) = 1$ for all $\Delta\chi^2_t$. One can even generalize the discussion to continuous priors $\pi(\theta \mid \Delta\chi^2_t)$, where in order to preserve the arguments on efficiency reduction, we would need to extend the single-point condition from Eq.~\ref{eq:mixture:deriv:hatcond} to a condition on a finite-size region on $\pi(\theta \mid \Delta\chi^2_t)$.

Unlike in the conventional FC method, where one needs a large number of pseudo-experiments at each target parameter value, it can be preferable in the mixture-FC method to generate less pseudo-experiments at each sampling value, but instead increase the number of considered sampling points $S$. If $S n_\mathrm{exp}$ is held fixed, this results in a reduction of the variance of critical values by reducing the variance in weights bounded from above by $\exp(\epsilon/2)$.

Given this relation to the marginal distribution, let us now consider the computation of $\Delta\chi^2(x \mid \theta_t) = -2\log L(x \mid \theta_t) / L(x \mid \hat\theta(x))$ as being approximated by $-2\log L(x \mid \theta_t) / L_m(x)$, where in the denominator, the profiling operation was replaced by a marginalization over $\theta$ with some prior over $\theta$. We have therefore a simple likelihood ratio test between $p(x \mid \theta_t)$ and $p_m(x) \coloneq \int \mathrm d\theta \, \pi(\theta) p(x \mid \theta)$ and it now becomes evident that in order to efficiently generate pseudo-experiments with small $p$-values under the  null hypothesis $p(x \mid \theta_t)$, one should simply generate the pseudo-experiments from the alternative hypothesis $p_m(x)$, which is what is being done in the mixture-FC method. 

In practice, it will be easier to use the discrete ``prior'' over $\{\theta\}_S$ as was discussed in the text, because unless the likelihood is gaussian, the numerical integration required for marginalization usually increases the computational cost and complexity. This relation to the profiling/marginalization similarities can nevertheless be exploited to motivate an ideal spacing of $\{\theta\}_S$ values. Out of the well-known objective priors, Jeffreys' prior~\cite{doi:10.1098/rspa.1946.0056} is known to produce a prior that would be uniform in the parameterization in which the likelihood is gaussian, if such a parameterization exists. Since profiling and marginalization with a uniform prior over a gaussian likelihood produce equivalent results up to a constant offset, Jeffreys' prior can be considered a good candidate for choosing the $\{\theta\}_S$ values at which to generate pseudo-experiments. For example, in the CP-violation analysis that was discussed in the earlier section, it would be more suitable to choose a uniform spacing of parameter values not in $\delta_\mathrm{CP}$ but in $\sin\delta_\mathrm{CP}$ with equal probabilities for the sign of $\cos\delta_\mathrm{CP}$, since the dominant constraint is due to the total number of events $N \sim \mathrm{Poisson}(\lambda = A + B\sin\delta_\mathrm{CP})$ for some constants $A$ and $B$, resulting in an approximately gaussian likelihood over $\sin\delta_\mathrm{CP}$.

\subsection{Nuisance parameters}

Because the significant error-reduction in the mixture-FC method exploits the specific relation of the $\Delta\chi^2(\theta_t \mid x)$ statistic to the distribution that generates the pseudo-experiments,
one cannot assume all features to directly translate to an analysis with nuisance parameters or ``systematic'' parameters as they are often called in physics.
Especially for the commonly used methods of profile-FC~\cite{NOvA:2022wnj} or posterior Highland-Cousins methods~\cite{Cousins:1991qz}, where the
space of nuisance parameters from which to generate the pseudo-experiments is significantly reduced based
on constraints by the observed experimental data, it is possible to have situations where the straightforward application of the mixture-FC method does not yield the exponential reduction of errors on the estimated critical values given by Eq.~\ref{eq:mixture:perf:gamma}. One should therefore not rely on these to estimate the number of required pseudo-experiments.

In a relatively general setting, when the target distribution is directly a part of the mixture distribution (so $C=1$), one can show that even in the worst case, the variance on the critical values only increases very slightly compared to the conventional method, by a factor $1 / (1 - P(y))$ (see Appendix~\ref{sec:appendix:general-mix}). This factor is negligible considering that for high CL we have $P(y) \ll 1$. The weights are bounded from above by a similar limit, which is important for well defined importance sampling behavior. The naive application of the mixture-FC method to Feldman-Cousins confidence intervals is therefore still worth a try. In fact, certain situations may yield near-exponential reduction of errors as in the case without nuisance parameters, but due the lack of theoretical guarantees it is suggested to carefully study the distribution of weights and the reliability of bootstrap error estimates in this situation.

Because one cannot guarantee an exponential reduction of errors in a setting with nuisance parameters, the ability to interpolate critical values will be more interesting in this setting.
Here it is important that the pseudo-experiments generated between neighboring $\theta_s$ values (and suitable values of nuisance parameters) sufficiently overlap in the space of pseudo-experiments. Otherwise, the mismatch between pseudo-experiment generation and the statistical model behind the test statistic may quickly result in a large spread of weight values, that would make both the estimated critical values as well as their error estimates unreliable. This is because with nuisance parameters, there are significantly more dimensions in which the pseudo-experiments can differ, even if they have similar values for the test statistic.

In one specific situation however, all properties discussed in earlier sections are directly applicable despite the presence of nuisance parameters. This is when using the prior Highland-Cousins method in conjunction with a marginal-$\Delta\chi^2$ statistic, where it is essential to use the same prior distribution $\pi(\eta)$ for the nuisance parameters $\eta$ in both cases. This is because here the effect of nuisance parameters is entirely absorbed by the probability model to generate the pseudo-experiments, in the sense of $p(x \mid \theta) = \int \mathrm d\eta \, \pi(\eta) \, p(x \mid \theta,\eta)$, such that as far as the mixture-FC method is concerned, no nuisance parameters exist.

More detailed discussions with examples and possible modifications to the sampling distributions for pseudo-experiments will be discussed in a separate publication.

\subsection{Relation to similar techniques for statistical inference}

Very similar importance sampling techniques have been used for the calculation of $p$-values under a null hypothesis with a likelihood ratio statistic. For example, Woodroofe~\cite{WoodroofeB10} discusses the case with a continuous prior over the parameter of interest. In our notation,
\begin{equation}
	p_\mathrm{sample}(x) = \int \mathrm d\theta \, \pi(\theta) p(x \mid \theta)
\end{equation}
with only a lower bound on the weights
\begin{equation}
	w(x \mid \theta_t)
	\coloneq
	\frac{p(x \mid \theta_t)}{p_\mathrm{sample}(x)}
	\ge 
	\frac{p(x \mid \theta_t)}{p(x \mid \hat \theta(x))}
	= \exp\bigl[-\tfrac{1}{2} \Delta \chi^2(\theta_t \mid x)\bigr]
\end{equation}
given, rather than an upper bound which would be essential for showing small errors on the estimated $p$-values. An asymptotic formula for the weights using the Saddle-point method is also given.

Ref.~\cite{Cranmer:2014lly} (Sect.~5.6) describes a method developed in the search for the Higgs boson by the ATLAS experiment~\cite{ATLAS:2012yve}.
They point out the difficulty of performing the integral over the continuous prior in Woodroofe's method and instead use a set of discrete points $\{\theta\}_S$ including $\theta_t$, as we used for the mixture-FC method (with $C=1$).
The choice of weight function however is different in that a pseudo-experiment is used only if the 
\begin{equation}
	\omega_s(x \mid \theta_t)
	\coloneq
	\frac{p(x \mid \theta_t)}{p(x \mid \theta_s)}
	=
	\exp\bigl[ -\tfrac{1}{2} \left( \Delta \chi^2(\theta_t \mid x) - \Delta \chi^2(\theta_s \mid x)  \right)  \bigr]
\end{equation}
value at the parameter value $\theta_s$ from which the pseudo-experiment was sampled from is the smallest among all other values in $\{\theta\}_S$ --- i.e.\ $\omega_s(x \mid \theta_t) = \min_{s'} \omega_{s'}(x \mid \theta_t)$ --- and discarded otherwise. If the pseudo-experiment is used, it is weighted by $\omega_s(x \mid \theta_t)$.
Then by combining the pseudo-experiments sampled from all $\{\theta\}_S$ values with their weights, the desired distribution $p(x \mid \theta_t)$ is attained with higher probability to sample pseudo-experiments of large $\Delta \chi^2(\theta_t \mid x)$.
Since $\theta_t \in \{\theta\}_S$, this procedure ensures that $w(x \mid \theta_t) \le 1$ for well-behaved weights.

One downside of this vetoing technique, as explained by the authors, is that the spacing of $\{\theta\}_S$
must not be too dense in order not to reduce the efficiency of the method with a high vetoing probability.
The mixture-FC method does not have this problem, because the weights are computed using the actual sampling probability, which is the sum of probabilities over $\{\theta\}_S$, and no vetoing is necessary. 
While the claimed benefit of the vetoing technique is its independence from the exact normalization of the sampling probability distribution --- due to only using the probability \emph{ratios} $\omega_s$ --- the same is true for the choice of weights in the mixture-FC method, whose weights from Eq.~\eqref{eq:mixture:def:weight} can be written as
\begin{equation}
	w(x \mid \theta_s)
	= \frac{1}{\frac{1}{S} \sum_{s=1}^S \bigl[\omega_s(x \mid \theta_t)\bigr]^{-1}}
	.
\end{equation}

For the problem of finding $p$-values under a null hypothesis with a likelihood-ratio statistic, the relevant part of the mixture-FC can therefore be regarded a slight improvement to the method by Ref.~\cite{Cranmer:2014lly}. Furthermore, we have explicitly shown that under suitable conditions,
which for a typical setup requires the absence of nuisance parameters,
the variance on the estimated $p$-values is reduced exponentially for large values of the test statistic.

Finally, we note some of the differences of computing $p$-values to the FC confidence interval construction in the context of importance sampling. When computing $p$-values, we are typically interested in the distribution of the test statistic under a \emph{single} null hypothesis.
In contrast, in the FC method we need the test statistic distribution for \emph{all} plausible parameter values, which in practice is achieved by computing them for a finite set $\{\theta\}_S$, and interpolating in between.
The FC construction therefore benefits from the ability to interpolate critical values with importance sampling, which is not always of interest in the computation of $p$-values.
In addition, the pseudo-experiments sampled from different parameter values as required for the construction of the mixture distribution are already available even in the conventional FC method, making the transition to the mixture-FC method straightforward.

\section{Summary}

We presented a new method to compute critical values for Feldman-Cousins confidence intervals. 
The method is a simple extension of the conventional method in that the same sets of pseudo-experiments generated at different parameter values are simply combined with suitable weights.
We showed that this results in a significant reduction of the errors on the critical values, with exponential reduction for high confidence level critical values, at almost no additional computational cost.
The method was further shown to enable accurate interpolation of critical values between the parameter values at which the pseudo-experiments were generated.
The theoretically calculated performance was confirmed using a simple example for the analysis of neutrino oscillations.
While the exponential reduction of errors is currently only guaranteed for analyses without nuisance parameters,
the general technique is applicable to any analysis making use of the Feldman-Cousins method.

\appendix*
\section{Analysis of critical value variances for generic mixtures}
\label{sec:appendix:general-mix}

Let us denote the target distribution of pseudo-experiments at $\theta_t$ by $p_t(x)$. In a setting with nuisance parameters $\eta$ with probability distribution $p(x \mid \theta,\eta)$,
this could for example be $p\big(x \mid \theta_t, \dblhat{\eta}(\theta_t \mid x_\mathrm{obs})\big)$ for the profile-FC method, or $\int \mathrm d\eta \, \pi(\eta \mid x_\mathrm{obs}, \theta_t) \, p(x \mid \theta_t,\eta)$ in the posterior HC method, with $\dblhat{\eta}(\theta \mid x_\mathrm{obs}) = \argminNL_\eta \chi^2(\theta,\eta \mid x_\mathrm{obs})$ the profile best-fit values and $\pi(\eta \mid x_\mathrm{obs}, \theta)$ the posterior distribution for nuisance parameters conditioned by the target $\theta$ value for a fit to the observed data $x_\mathrm{obs}$. The other pseudo-experiments are sampled from $p_a(x)$, whose distribution we don't explicitly specify here, but could for example be a mixture over different $\theta$ and $\eta$ values. The mixture of $N_t$ pseudo-experiments sampled from $p_t(x)$ and $N_a$ pseudo-experiments sampled from $p_a(x)$ weighted by $w(x) = (N_t + N_a) p_t(x) / \big[N_t p_t(x) + N_a p_a(x)\big]$ can be evaluated analogously to the main text and using the estimators 
\begin{align}
	\hat P(y)
	&\coloneq \frac{1}{N_t + N_a} \sum_{i = 1}^{N_t + N_a} w(x_i) I\big( Y(x_i) \ge y \big)
	\\
	\hat P_\mathrm{conv}(y)
	&\coloneq \frac{1}{N_t} \sum_{i = 1}^{N_t} I\big( Y(x_i^{(t)}) \ge y \big)
	\\
	\EE[\hat P(y)] &= \EE[\hat P_\mathrm{conv}(y)] = P(y)
	\\
\intertext{yield a variance reduction of}
	\gamma 
	&= \frac{\Var[\hat P(y)]}{\Var[\hat P_\mathrm{conv}(y)]} \\
	&\le \frac{\frac{1}{N_t} P(y) - \frac{1}{N_t + N_a} P(y)^2}{\frac{1}{N_t} P(y) - \frac{1}{N_t} P(y)^2} \\
	&\le \frac{1}{1 - P(y)}
	\intertext{where}
	x_i &\sim \frac{N_t p_t(x) + N_a p_a(x)}{N_t + N_a}
	\\
	x_i^{(t)} &\sim p_t(x)
	.
\end{align}

\begin{acknowledgments}
We would like to thank Christophe Bronner and Louis Lyons for useful discussions and connecting us to Kyle Cranmer, whom we would like to thank for introducing Refs.~\cite{WoodroofeB10,Cranmer:2014lly}. This research was supported by JSPS KAKENHI Grant Number 19J22440.
\end{acknowledgments}

\bibliography{mixture-FC}

\end{document}